\documentclass[apj,numberedappendix]{emulateapj-rtx4}
\usepackage{amssymb,amsmath}
\usepackage{graphicx}
\usepackage{appendix}
\usepackage{natbib}
\usepackage{hyperref}
\usepackage{enumerate}
\usepackage{multirow}
\usepackage{bm}
\usepackage{url}

\begin{document}


\title{Deep Chandra, HST-COS, and Megacam Observations of the Phoenix Cluster: \\Extreme Star Formation and AGN Feedback on Hundred Kiloparsec Scales}

\email{mcdonald@space.mit.edu}



\def\MIT{1}
\def\Waterloo{2}
\def\Perimeter{3}
\def\CfA{4}
\def\AlfA{5}
\def\Harvard{6}
\def\FNAL{7}
\def\AAC{8}
\def\KICP{9}
\def\ANL{10}
\def\Kentucky{11}
\def\Durham{12}
\def\Cambridge{13}
\def\Huntingdon{14}
\def\Montreal{15}
\def\UChicago{16}
\def\IfA{17}
\def\UMd{18}
\def\JSI{19}

\altaffiltext{\MIT}{Kavli Institute for Astrophysics and Space Research, MIT, Cambridge, MA 02139, USA}
\altaffiltext{\Waterloo}{Department of Physics and Astronomy, University of Waterloo, Waterloo, ON N2L 3G1, Canada}
\altaffiltext{\Perimeter}{Perimeter Institute for Theoretical Physics, Waterloo, Canada}
\altaffiltext{\CfA}{Harvard-Smithsonian Center for Astrophysics, 60 Garden Street, Cambridge, MA 02138, USA}
\altaffiltext{\AlfA}{Argelander-Institut f{\"u}r Astronomie, Auf dem H{\"u}gel 71, D-53121 Bonn, Germany}
\altaffiltext{\Harvard}{Department of Physics, Harvard University, 17 Oxford Street, Cambridge, MA 02138}
\altaffiltext{\FNAL}{Fermi National Accelerator Laboratory, Batavia, IL 60510-0500, USA}
\altaffiltext{\AAC}{Department of Astronomy and Astrophysics, University of Chicago, 5640 South Ellis Avenue, Chicago, IL 60637, USA}
\altaffiltext{\KICP}{Kavli Institute for Cosmological Physics, University of Chicago, 5640 South Ellis Avenue, Chicago, IL 60637, USA}

\altaffiltext{\ANL}{Argonne National Laboratory, High-Energy Physics Division, 9700 South Cass Avenue, Argonne, IL 60439, USA}
\altaffiltext{\Kentucky}{Department of Physics \& Astronomy, University of Kentucky, Lexington, KY 40506, USA}
\altaffiltext{\Durham}{Department of Physics, Durham University, Durham DH1 3LE, UK}
\altaffiltext{\Cambridge}{Institute of Astronomy, Madingley Road, Cambridge CB3 0HA, UK}
\altaffiltext{\Huntingdon}{Huntingdon Institute for X-ray Astronomy, LLC}
\altaffiltext{\Montreal}{D\'{e}partement de Physique, Universit\'{e} de Montr\'{e}al, C.P. 6 128, Succ. Centre-Ville, Montreal, Quebec H3C 3J7, Canada}
\altaffiltext{\UChicago}{Department of Astronomy and Astrophysics, University of Chicago, 5640 South Ellis Avenue, Chicago, IL 60637, USA}
\altaffiltext{\IfA}{Institute for Astronomy, University of Hawaii, 2680 Woodlawn Drive, Honolulu, HI 96822, USA}
\altaffiltext{\UMd}{Department of Astronomy, University of Maryland, College Park, MD 20742, USA}
\altaffiltext{\JSI}{Joint Space-Science Institute, University of Maryland, College Park, MD 20742, USA}

\author{
Michael McDonald$^{\MIT}$,
Brian R.\ McNamara$^{\Waterloo,\Perimeter}$,
Reinout J.\ van Weeren$^{\CfA}$,
Douglas E.\ Applegate$^{\AlfA}$,\\
Matthew Bayliss$^{\CfA,\Harvard}$,
Marshall W.\ Bautz$^{\MIT}$,
Bradford A.\ Benson$^{\FNAL,\AAC,\KICP}$,
John E.\ Carlstrom$^{\AAC,\KICP,\ANL}$,\\
Lindsey E.\ Bleem$^{\KICP,\ANL}$,
Marios Chatzikos$^{\Kentucky}$,
Alastair C.\ Edge$^{\Durham}$,
Andrew C.\ Fabian$^{\Cambridge}$,\\
Gordon P.\ Garmire$^{\Huntingdon}$,
Julie Hlavacek-Larrondo$^{\Montreal}$,
Christine Jones-Forman$^{\CfA}$,
Adam B.\ Mantz$^{\UChicago,\KICP}$,\\
Eric D.\ Miller$^{\MIT}$,
Brian Stalder$^{\IfA}$,
Sylvain Veilleux$^{\UMd,\JSI}$,
John A.\ ZuHone$^{\MIT}$
}



\begin{abstract}
We present new ultraviolet, optical, and X-ray data on the Phoenix galaxy cluster (SPT-CLJ2344-4243). Deep optical imaging reveals previously-undetected filaments of star formation, extending to radii of $\sim$50--100~kpc in multiple directions. Combined UV-optical spectroscopy of the central galaxy reveals a massive ($2\times10^9$~M$_{\odot}$), young ($\sim$4.5~Myr) population of stars, consistent with a time-averaged star formation rate of $610 \pm 50$~M$_{\odot}$~yr$^{-1}$. We report a strong detection of O\,\textsc{vi} $\lambda\lambda$1032,1038, which appears to originate primarily in shock-heated gas, but may contain a substantial contribution ($>$1000 M$_{\odot}$ yr$^{-1}$) from the cooling intracluster medium. We confirm the presence of deep X-ray cavities in the inner $\sim$10 kpc, which are amongst the most extreme examples of radio-mode feedback detected to date, implying jet powers of $2-7\times10^{45}$ erg s$^{-1}$. We provide evidence that the AGN inflating these cavities may have only recently transitioned from ``quasar-mode'' to ``radio-mode'', and may currently be insufficient to completely offset cooling. A model-subtracted residual X-ray image reveals evidence for prior episodes of strong radio-mode feedback at radii of $\sim$100~kpc, with extended ``ghost'' cavities indicating a prior epoch of feedback roughly 100 Myr ago. This residual image also exhibits significant asymmetry in the inner $\sim$200~kpc (0.15R$_{500}$), reminiscent of infalling cool clouds, either due to minor mergers or fragmentation of the cooling ICM. Taken together, these data reveal a rapidly evolving cool core which is rich with structure (both spatially and in temperature), is subject to a variety of highly energetic processes, and yet is cooling rapidly and forming stars along thin, narrow filaments.
\end{abstract}


\section{Introduction}
\setcounter{footnote}{0}

The hot intracluster medium (ICM) is the most massive baryonic component in galaxy clusters, comprising $\sim$12\% of the total cluster mass, or $\sim$10$^{13}$--10$^{14}$ M$_{\odot}$ in rich clusters. The bulk of the ICM is very low density ($\ll10^{-3}$ cm$^{-3}$), and will require $\sim$10 Gyr to cool via thermal Bremsstrahlung radiation. In the centers of clusters, however, this situation is reversed. The ICM in the inner $\sim$100 kpc can reach high enough density that the cooling time is short relative to the age of the cluster ($t_{cool} \lesssim 1$ Gyr). This ought to result in a runaway cooling flow, fueling a massive, 100--1000 M$_{\odot}$ yr$^{-1}$ starburst in the central cluster galaxy \citep[for a review, see][]{fabian94}. However, despite the fact that roughly a third of all galaxy clusters have short central cooling times \citep[e.g.,][]{bauer05,vikhlinin07,hudson10,mcdonald11c, mcdonald13b}, such massive starbursts are extremely rare \citep[e.g.,][]{mcnamara06,mcdonald12c}. The vast majority of clusters which ought to host runaway cooling flows show only mild amounts of star formation \citep{johnstone87,mcnamara89,allen95,hicks05,odea08,mcdonald11b,hoffer12,donahue15,mittal15}, suggesting that much of the predicted cooling at high temperature is not, in fact, occurring.

By comparing the cooling rate of intermediate-temperature ($\sim$10$^{5-6}$ K) gas \citep[e.g.,][]{bregman01,oegerle01,peterson03,bregman06,peterson06,sanders10,sanders11,mcdonald14b} to the ongoing star formation rate in central cluster galaxies, one can quantify what fraction of the cooling ICM is able to condense and form stars. These studies typically find that only $\sim$1\% of the predicted cooling flow is ultimately converted into stars. Recently, \cite{mcdonald14b} showed that this inefficient cooling can be divided into two contributing parts: cooling from high ($\sim$10$^7$ K) to low ($\sim$10 K) temperature at $\sim$10\% efficiency, and an additionally $\sim$10\% efficiency of converting cold gas into stars. The former term is generally referred to as the ``cooling flow problem'', and suggests that some form of feedback is preventing gas from cooling out of the hot phase.

The most popular solution to the cooling flow problem is that ``radio-mode'' feedback from the central supermassive blackhole is offsetting radiative losses in the ICM \citep[see reviews by][]{mcnamara12,fabian12}. Radio jets from the active galactic nucleus (AGN) in the central cluster galaxy can inflate bubbles in the dense ICM, imparting mechanical energy to the hot gas \citep[e.g.,][]{birzan04,birzan08,dunn05,dunn06,rafferty06,nulsen07,cavagnolo10,dong10,osullivan11,hlavacek12,hlavacek14}. The ubiquity of radio jets in so-called ``cool core clusters'' \citep[e.g.,][]{sun09b} suggests that the two are intimately linked, while the correlation of the feedback strength (e.g., radio power, mechanical energy in bubbles) with the X-ray cooling luminosity provides evidence that this mode of feedback is, on average, sufficient to fully offset cooling \citep[e.g.,][]{rafferty06,hlavacek12}. Further, recent studies of newly-discovered high-$z$ clusters in the South Pole Telescope 2500 deg$^2$ survey \citep{bleem15} suggest that this energy balance between ICM cooling and AGN feedback has been in place since $z\sim1$ \citep{mcdonald13b,hlavacek14}.

The study of outliers in a population can often provide new insights into the physical processes governing said population. Amongst cool core clusters, there is perhaps no more extreme system than the Phoenix cluster \citep[SPT-CLJ2344-4243;][]{williamson11,mcdonald12c}. This massive ($M_{500} \sim 1.3\times10^{15}$ M$_{\odot}$) system is the most X-ray luminous cluster yet discovered (L$_{2-10\,\textrm{keV},500} = 8.2 \times 10^{45}$ erg s$^{-1}$), with a predicted cooling rate of $\sim$2000 M$_{\odot}$ yr$^{-1}$. However, contrary to the norm, the central galaxy in the Phoenix cluster harbors a massive starburst \citep[$\sim$800 M$_{\odot}$ yr$^{-1}$;][]{mcdonald13a}, a massive reservoir of molecular gas \citep[M$_{\textrm{H}_2} \sim 2\times10^{10}$ M$_{\odot}$;][]{mcdonald14a}, and a dusty type-2 quasar \citep[QSO;][]{mcdonald12c,ueda13}. In \cite{mcdonald12c}, we argued that the only feasible way to bring such a vast supply of cold gas into the center of a rich cluster was via a runaway cooling flow. However, \cite{hlavacek14} showed that radio-mode feedback is operating at a level that ought to offset radiative cooling in this system, although whether this energy has had time to couple to the ICM is uncertain. The fact that the central AGN is simultaneously providing strong radiative (quasar-mode) and mechanical (radio-mode) feedback suggests that it may be in the process of transitioning from a QSO to a radio galaxy, and that the starburst is being fueled by gas that cooled before the first radio outburst. However, this interpretation hinges on a single, shallow X-ray observation with the \emph{Chandra X-ray Observatory} (10 ks), in which both the X-ray nucleus and cavities are detected at low significance.

In an effort to provide a more complete picture of this system, we have obtained deep far-UV spectroscopy and X-ray imaging spectroscopy using the \emph{Hubble Space Telescope} Cosmic Origins Spectrograph (HST-COS) and the Advanced CCD Imaging Spectrometer (ACIS) on the \emph{Chandra X-ray Observatory}, respectively. These data provide a detailed picture of the young stellar populations, the intermediate-temperature gas (O\textsc{vi}; 10$^{5.5}$K), and the hot intracluster medium. These new X-ray data represent a factor of $>$10 increase in exposure time over previously-published observations. We have also obtained deep optical and radio data via Megacam on the Magellan Clay Telescope and the Giant Meterwave Radio Telescope (GMRT), respectively. Combined, these data will provide a much more detailed view of the complex interplay between cooling and feedback in this extreme system. We describe the reduction and analysis of these new data in \S2, along with supporting data at IR and optical wavelengths. In \S3 we present new results derived from these data, focusing on the central AGN, the starburst, and the cluster core. In \S4 we discuss the implications of these results and present a coherent picture of the physical processes at work in this unique system. We conclude in \S5 with a summary of this and previous work, with a look towards the future.

Throughout this work, we assume H$_0 = 70$ km s$^{-1}$ Mpc$^{-1}$, $\Omega_M = 0.27$, and $\Omega_\Lambda = 0.73$. We assume $z=0.596$ for the Phoenix cluster, which is based on optical spectroscopy of the member galaxies \citep{ruel14,bleem15}.

\section{Data}
Below, we summarize the acquisition and reduction of new data used in this study, along with a brief summary of supporting data published in previous works.

\subsection{Chandra: X-ray Imaging Spectroscopy}
\begin{figure}[htb]
\centering
\includegraphics[width=0.47\textwidth]{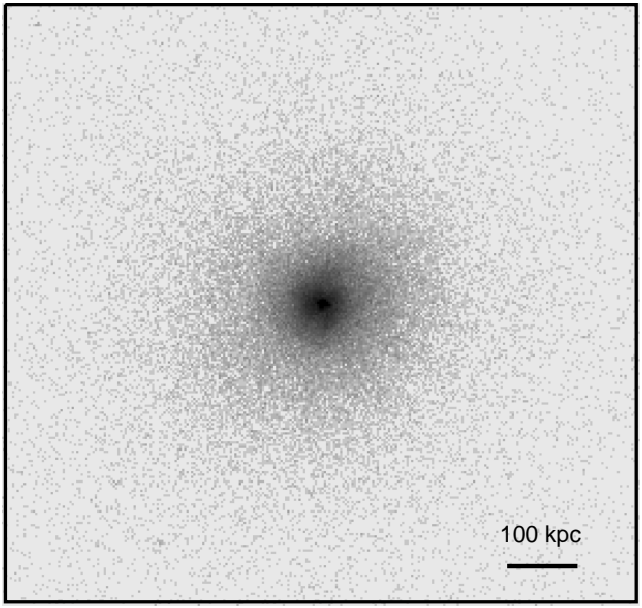}
\caption{\emph{Chandra} ACIS-I 0.5--8.0 keV image of the Phoenix cluster, representing a total exposure of 129.3\.ks. This deep exposure reveals a relatively relaxed morphology with the majority of the counts being concentrated in the central $\sim$100\,kpc. }
\label{fig:chandra}
\end{figure}

The Phoenix cluster (SPT-CLJ2344-4243) was observed by \emph{Chandra} in 2011 (\textsc{obsid}: 13401; PI: Garmire) for a total of 11.9\,ks as part of a large X-ray survey of SPT-selected clusters (PI: Benson). To understand in detail this extreme system, we have obtained an additional 117.4\,ks (\textsc{obsid}s: 16135, 16545; PIs: McDonald, Garmire), resulting in a combined exposure time of 129.3\,ks (see Figure \ref{fig:chandra}) and a total of 88,042 counts in the central 1\,Mpc and in the energy range 0.5--8.0 keV. All observations were performed with ACIS-I. The data from each observation were individually processed in the standard manner, including cleaning and filtering for background flares, using CIAO 4.6 and CALDB 4.6.3. 
The X-ray background was modeled using three distinct components. First, the re-scaled ACIS blank-sky background observations provide an adequate representation of both the particle-induced and unresolved cosmic X-ray background. A second component, modeled as a soft (0.18\,keV \textsc{apec}) excess, accounts for Galactic interstellar medium emission \citep{markevitch03}. Finally, a third component, modeled as a hard (absorbed 40\,keV \textsc{bremss}) excess, accounts for the fact that a larger-than-normal fraction of the CXB may be unresolved in shorter exposures. The latter two components were simultaneously fit to the on-source spectrum and an off-source spectrum extracted from a blank region of the target field, at a physical separation of $>$3\,Mpc from the cluster center.

In order to search for structure in the X-ray surface brightness, images were made in several bandpasses including a soft (0.5--2.0 keV) and hard (4.0--8.0 keV) band, which trace the hot ICM and central AGN, respectively. To search for faint structure in the soft image, we subtract a two-dimensional model comprised of three beta functions with a shared center, position angle, and ellipticity. The residual image after this model was subtracted was smoothed both adaptively, using \textsc{csmooth}\footnote{\url{http://cxc.harvard.edu/ciao/ahelp/csmooth.html}}, and with fixed-width Gaussians. The former technique provides the highest-quality image over large scales, while the latter has a lower (but non-zero) likelihood of artificially creating features out of noise.

Thermodynamic profiles were measured by extracting on-source spectra in concentric annuli, where we strive for $\sim$5000 X-ray counts per annulus. These spectra were empirically deprojected using \textsc{dsdeproj} \citep{sanders07,russell08}. Deprojected spectra were background subtracted using off-source regions (the cluster only occupies one of the four ACIS-I chips) and fit over the range 0.5--10.0\,keV using \textsc{xspec} \citep{arnaud96}. We model the X-ray spectrum with a combination of Galactic absorption (\textsc{phabs}) and an optically-thin plasma (\textsc{mekal}), freezing the absorbing column ($n_H$) to the Galactic value \citep{kalberla05} and the redshift to $z=0.596$. In the inner two annuli, we add an additional absorbed powerlaw component to account for emission from the central X-ray-bright AGN.

\subsection{HST-COS: Far UV Spectroscopy}

\begin{figure*}[htb]
\includegraphics[width=0.98\textwidth]{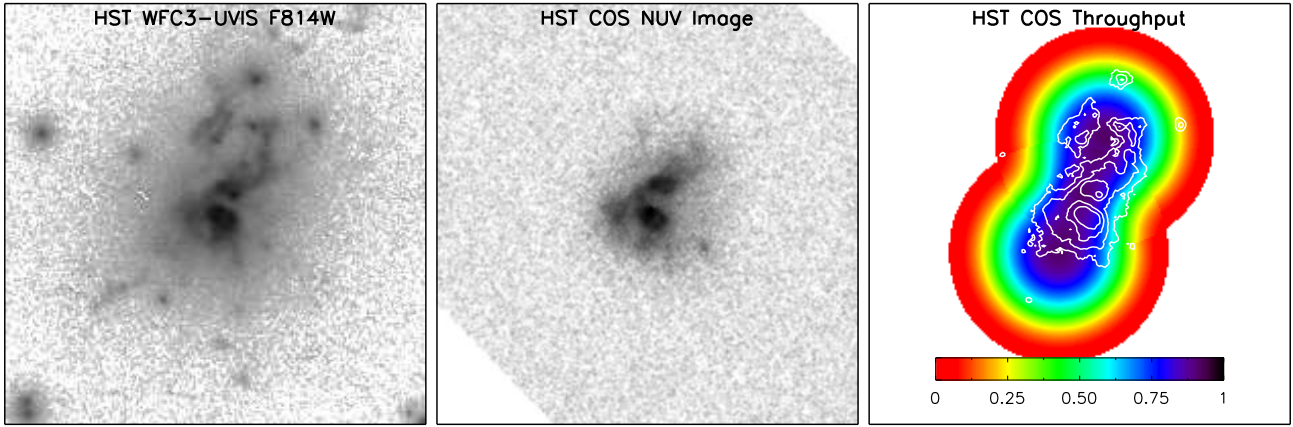}
\caption{Left panel: HST WFC3-UVIS image of the central galaxy in the Phoenix cluster, from \cite{mcdonald13a}. Middle panel: HST-COS near-UV image on the same scale, showing that the UV continuum comes from scales larger than the central AGN. Right panel: Combined spectroscopic throughput from two pointings of HST-COS. White contours show the HST WFC3-UVIS  $i$-band imaging, which is well covered by the COS apertures.}
\label{fig:cos_regions}
\end{figure*}

The Cosmic Origins Spectrograph (COS) is a medium-to-high resolution UV spectrograph on HST with a 2.5$^{\prime\prime}$ diameter aperture. 
The broad UV coverage and sensitivity of this instrument make it ideal for simultaneously detecting coronal line emission (e.g., O\,\textsc{vi}) and the stellar continuum in young stellar populations.
We designed our observational setup (ID 13456, PI: McDonald) to be able to detect O\,\textsc{vi}\,$\lambda\lambda$1032,1038 over the full extent of the central starburst. This required two offset pointings, as depicted in Figure \ref{fig:cos_regions}. Following the COS Instrument Handbook\footnote{\url{http://www.stsci.edu/hst/cos/documents/handbooks/current/cos_cover.html}}, we model the throughput across the aperture, yielding a two-dimensional estimate of the throughput over the extent of the central cluster galaxy. Our setup yields $>$75\% throughput over the full extent of the starburst, with $>$50\% throughput out to the edge of the central galaxy. Figure \ref{fig:cos_regions} shows our estimate of the two-dimensional throughput map compared to optical and near-UV images from HST WFC3-UVIS and COS of the central cluster galaxy.

We use the G160M grating with a central wavelength of 1577\AA, providing coverage from 1386--1751\AA. This corresponds to rest-frame 870--1100\AA\ at $z=0.596$, which covers the Ly$\beta$, O\,\textsc{vi}\,$\lambda\lambda$1032,1038, and He\,\textsc{ii}\,$\lambda$1085 lines. Each of the two pointings were observed for 8 orbits, totaling $\sim$20\,ks on-source for each spectrum. This exposure time was chosen to allow simultaneous modeling of the O\,\textsc{vi} lines and the UV continuum, to sufficient depth to rule out a 1\% efficient cooling flow in the absence of an O\,\textsc{vi} detection.

The observed data were binned to $\sim$0.8\AA/pix to improve the signal-to-noise. Geocoronal emission and residual background was subtracted using a deep blank-sky spectrum with the same observational setup\footnote{http://www.stsci.edu/hst/cos/calibration/airglow.html}. The resulting spectra were modeled using the latest version of Starburst99 \citep[v7.0.0;][]{leitherer99}\footnote{http://www.stsci.edu/science/starburst99/docs/default.htm}. These models have sufficient spectral resolution ($<1$\AA) over 1000\AA~$< \lambda < $~1100\AA\ to constrain the age and metallicity of the starburst in the core of Phoenix. We are restricted to the empirically-calibrated models in this wavelength range, which have metallicities of $Z_{\odot}$ (Galactic) and $Z_{\odot}/7$ (LMC/SMC). We use the latest models from \cite{ekstrom12} and \cite{georgy13}, which include rotation of massive stars. The fits were performed by grid searching over a variety of populations (continuous/instantaneous star formation, low/high metallicity), with the ages ranging from 0--50 Myr in steps of 0.5 Myr. For each grid position, we perform a least-squares fit of the model to the data, allowing the normalization, reddening, redshift, stellar dispersion, and emission line fluxes (Ly$\beta$, O\,\textsc{vi}\,$\lambda\lambda$1032,1038, and He\,\textsc{ii}\,$\lambda$1085) to vary. The best-fitting model was chosen to minimize $\chi^2$. The best-fitting models are shown in Figures \ref{fig:fullspec} and \ref{fig:sb99model}, and will be discussed in \S3.1 and \S4.1.

\subsection{MegaCam Optical Imaging}

The Phoenix cluster (SPT-CLJ2344-4243) was observed on 30 August 2013 with the Megacam instrument on the Clay-Magellan telescope at Las Campanas Observatory, Chile \citep{mcleod1998}. Exposures were taken in $g$ (4$\times$200 sec), $r$ (4$\times$600 sec + 2$\times$120 sec), and $i$ (4$\times$400 sec) passbands in standard operating mode and cover the inner 12$^{\prime}$ (5\,Mpc) of the cluster. For this work we consider only the inner $\sim$0.5$^{\prime}$ -- the remaining data will be published in an upcoming weak lensing study (Applegate et al.\ in prep). Images were processed and stacked using the standard Megacam reduction pipeline at the Smithsonian Astrophysical Observatory (SAO) Telescope Data Center.  Color photometry was calibrated using Stellar Locus Regression \citep[SLR;][]{high2009}. For a more detailed description of the data reduction and calibration, the reader is directed to \cite{high2012}.

\subsection{Giant Meterwave Radio Telescope: 610 MHz}

Giant Metrewave Radio Telescope (GMRT) 610~MHz observations of SPT-CL~J2344--4243  were taken on June 14 and 15, 2013. The total on source time resulting from these two observing runs was about 10~hrs, with a usable bandwidth of 29~MHz. The data were reduced with the Astronomical Image Processing System\footnote{http://www.aips.nrao.edu} (AIPS), ParselTongue \citep{kettenis06} and Obit \citep{cotton08}, following the scheme detailed in \cite{intema09}. Reduction steps include flagging of  radio frequency interference (RFI), bandpass and gain calibration. Following that, several cycles of self-calibration were carried out to refine the calibration solutions. Direction-dependent gain solutions were then obtained towards several bright sources within the field of view. The data was imaged by dividing the field of view into smaller facets \citep{perley89,cornwell92}. This corrects for the non-coplanar nature of the array and the direction-dependent calibration.  For more details about the data reduction the reader is referred to \cite{intema09} and \cite{vanweeren14}.

\subsection{Supporting Data: UV, Optical, IR}
We also include in this paper optical multi-band imaging from HST WFC3-UVIS \citep{mcdonald13a} and optical imaging spectroscopy \citep{mcdonald14a}. For a detailed description of a specific dataset, we direct the reader to the aforementioned papers. Relevant features of the various datasets are summarized below.

Optical imaging was acquired on the cluster core for \cite{mcdonald13a} in five WFC3-UVIS bands: F225W, F336W, F475W, F625W, and F814W. While relatively shallow, these data provided the first resolved view of the starburst in the central cluster galaxy. Additionally, large-area ground-based data at $z$-band was obtained with MOSAIC-II on the Blanco 4-m telescope \citep[for details, see][]{song12,bleem15}.

Optical imaging spectroscopy of the central galaxy was obtained from Gemini-S GMOS in IFU mode, and presented in \cite{mcdonald14a}. These data span rest-frame 3000--6000\AA, and reveal complex emission-line nebulae in and around the central galaxy. The morphology of this gas is, for the most part, consistent with the near-UV morphology. To allow a direct comparison of HST-COS and GMOS spectroscopy, we extract aperture spectra from the GMOS IFU data, matching the COS throughput as a function of radius within the aperture. This allows us to simultaneously model the UV and optical continuum, as well as compare measured emission line fluxes in the UV and optical.


\section{Results}
Below, we summarize the main results that emerge from the analysis of these deep UV, X-ray, optical, and radio data. We defer a discussion of these results in the greater context of the cluster's evolution to \S4.

\subsection{Dust and Young Stellar Populations in the BCG}

 \begin{figure*}[htb]
 \centering
\includegraphics[width=0.99\textwidth, trim=0cm 0cm 1cm 0cm]{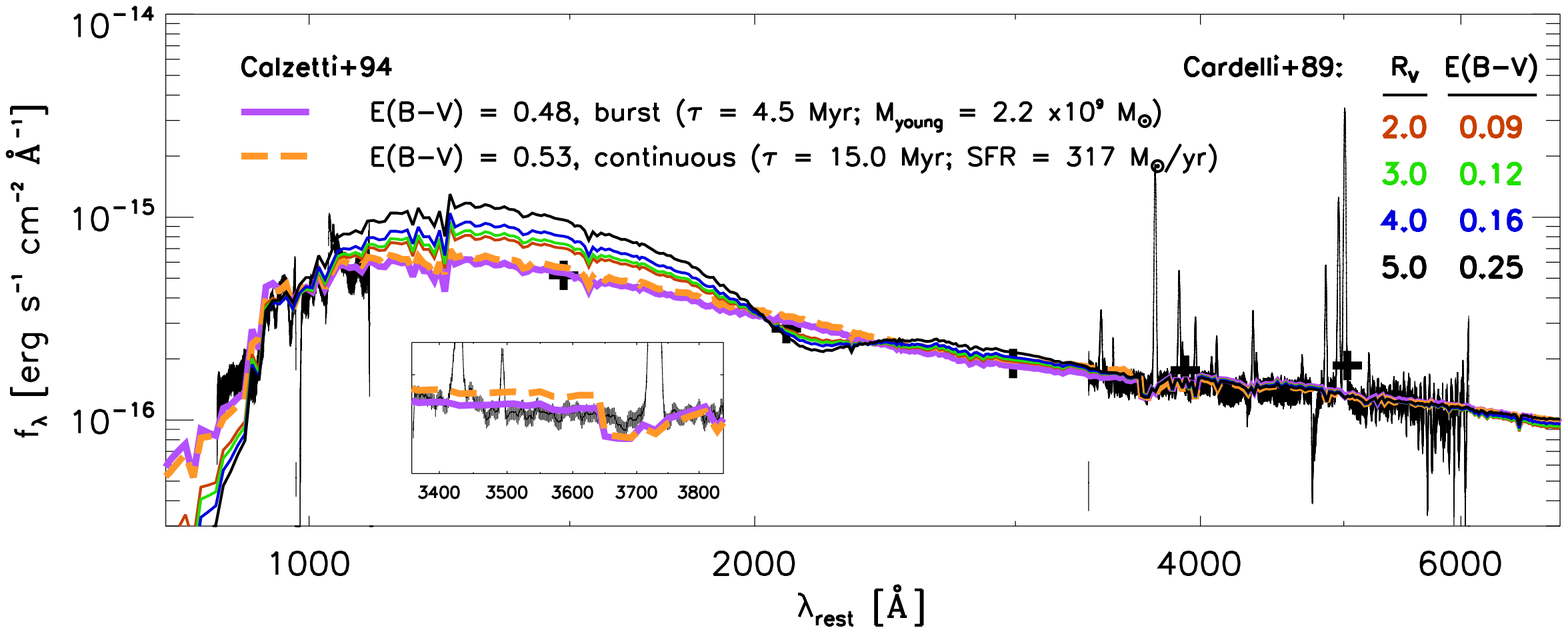}
\caption{UV-through-optical spectrum of the central galaxy in the Phoenix cluster. The optical IFU spectrum, from \cite{mcdonald14a}, and broadband photometry, from \cite{mcdonald13a}, have been aperture matched to the far-UV HST-COS spectrum based on Figure \ref{fig:cos_regions}. Thin colored curves show the best-fit stellar population model reddened by a variety of Galactic extinction models from \cite{cardelli89}, while thick orange and purple lines show the best-fit models using extinction models from \cite{calzetti94}. The featureless, grey extinction model from \cite{calzetti94} provides a better match to the spectrum, which appears to lack the characteristic 2175\AA\ absorption feature. The best-fit stellar population model consists of a highly-reddened, young (4.5 Myr), metal-poor starburst. A continuous starburst model provides a qualitatively similar fit, although the strength of the Balmer jump (inset) appears to be more consistent with a recently-quenched starburst.
}
\label{fig:fullspec}
\end{figure*}

\begin{figure}[htb]
\centering
\includegraphics[width=0.48\textwidth]{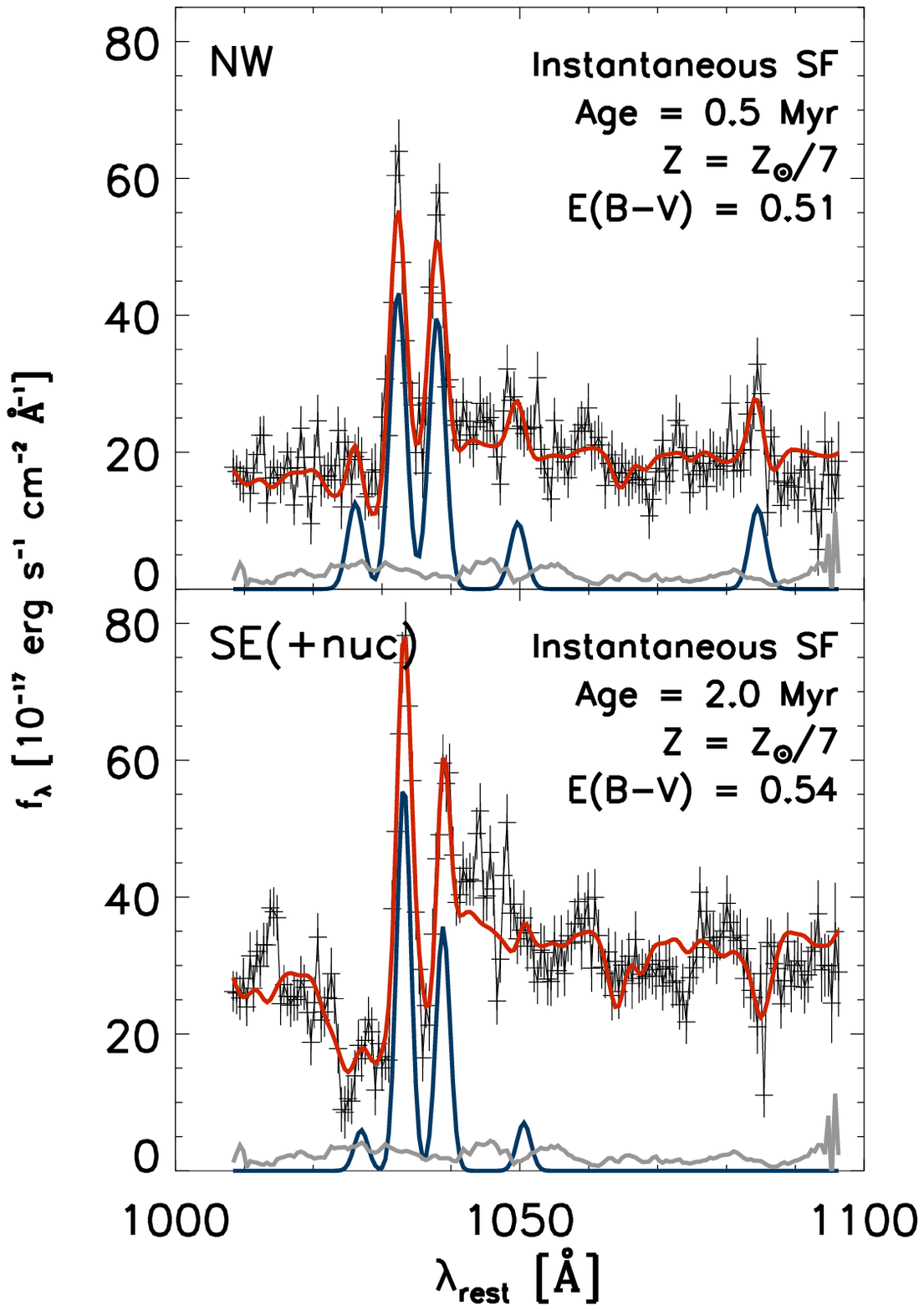}
\caption{Far-UV spectra for the two HST-COS pointings described in \S2.2. In each panel, the red curves shows the best-fit young stellar population model, including emission from Ly$\beta$, O\,\textsc{vi} $\lambda\lambda$1032,1038, [Si\,\textsc{vi}]\,$\lambda$1049, and He\,\textsc{ii}\,$\lambda$1085 (blue) and a blank-sky background component (grey). The best-fitting stellar population model is described in the upper right of each panel. In general, these spectra are consistent with a highly-reddened, young ($<$5 Myr) population of stars.}
\label{fig:sb99model}
\end{figure}

Using aperture-matched spectra from both HST-COS and Gemini-S GMOS \citep{mcdonald14a}, combined with broadband photometry spanning the spectral gap at 1200--3500\AA\ \citep{mcdonald13a}, we can constrain the relative contribution of young and old stellar populations to the UV+optical continuum, along with the effects of reddening on the spectrum over nearly an order of magnitude in wavelength.
In Figure \ref{fig:fullspec}, we show the full UV+optical spectrum for the combined area of the two HST-COS apertures (Figure \ref{fig:cos_regions}). We overplot a series of best-fitting model spectra, comparing different reddening models \citep{cardelli89,calzetti94} and star formation histories. All models include a variable-mass old (6 Gyr) stellar population. We note that the extinction curve from \cite{cardelli89} was calibrated using observations of stars in our galaxy and the Magellanic clouds, while \cite{calzetti94} was calibrated on nearby starburst and blue compact galaxies. We find, as did \cite{calzetti94}, that a relatively gray, ``feature-less'' extinction curve provides the best fit to the data, suggesting that the size distribution of the dust is skewed towards larger grains. The data also suggest a lack of the characteristic 2175\AA\ bump, which is the strongest absorption feature in the interstellar medium but tends to be missing in starburst galaxies \citep{calzetti94}. \cite{fischera11} suggest that high levels of turbulence can both flatten the curvature of the extinction law and wipe out the 2175\AA\ absorption feature. This is consistent with \cite{mcdonald14a}, where we show that the ionized gas has significant velocity structure ($\left<\sigma_v\right> \sim  300$ km s$^{-1}$) and is consistent with being ionized primarily by shocks.

Assuming the flatter extinction curve of \cite{calzetti94}, we find an overall good fit ($\chi^2_{dof} = 1.93$) to the data spanning rest-frame 800--6000\AA. The best-fitting model (Figure \ref{fig:fullspec}; purple curve) is a 4.5\,Myr-old starburst with a total zero-age mass of $2.2\times10^9$ M$_{\odot}$. \emph{This corresponds to $<$1\% of the total stellar mass of the central galaxy.} The fit quality is only slightly reduced ($\chi^2_{dof} = 1.97$) if we assume a continuous star formation history, with the best-fit model representing a $\sim$317 M$_{\odot}$ yr$^{-1}$ over the past $\gtrsim$15 Myr (Figure \ref{fig:fullspec}; orange curve). These two models disagree on the strength of the Balmer jump at $\sim$3650\AA\ (see inset of Figure \ref{fig:fullspec}), with the data preferring the instantaneous star formation model. This provides marginal evidence that star formation has been quenched in the past $\sim$5 Myr.

When the far-UV spectrum (1000\AA\ $<$ $\lambda_{rest}$ $<$ 1100\AA) is considered on its own, the data prefer a younger population ($\sim$1--2 Myr), as we show in Figure \ref{fig:sb99model}. It is worth noting that the age is only weakly constrained in this region of the spectrum, due to the relatively short wavelength range. There is little difference in stellar populations and reddening between the northern and southern COS apertures. This may be due to the considerable overlap between the two regions (see Figure \ref{fig:cos_regions}) or may indicate a uniformity in the starburst. With only two options for metallicity when synthesizing high-resolution spectra at far-UV \citep[][see also \S2.2]{ekstrom12,georgy13}, we can only weakly constrain the metallicity from this spectrum. We note that the solar metallicity spectrum provides a significantly worse fit than the spectrum for the $Z=Z_{\odot}/7$ population, with the former having significantly (factor of $\sim$5) stronger absorption lines which are not present in the observed spectrum. 
We stress that, despite the extremely high star formation rate inferred by the UV spectrum, the young population constitutes $<$1\% of the total stellar mass in the central galaxy. This is consistent with our picture of BCG formation, since this mode of star formation \emph{must} be suppressed to prevent central cluster galaxies from growing too massive (and luminous) by $z\sim0$.

In summary, the combined UV-through-optical spectrum of the central galaxy in the Phoenix cluster reveals a highly-reddened (E(B-V) $\sim$ 0.5), young stellar population, consistent with being recently quenched ($\lesssim$5 Myr).

\subsection{Coronal Emission: O\,\textsc{vi}\,$\lambda\lambda$1032,1038}

\begin{figure}[htb]
\centering
\begin{tabular}{c}
\includegraphics[width=0.48\textwidth]{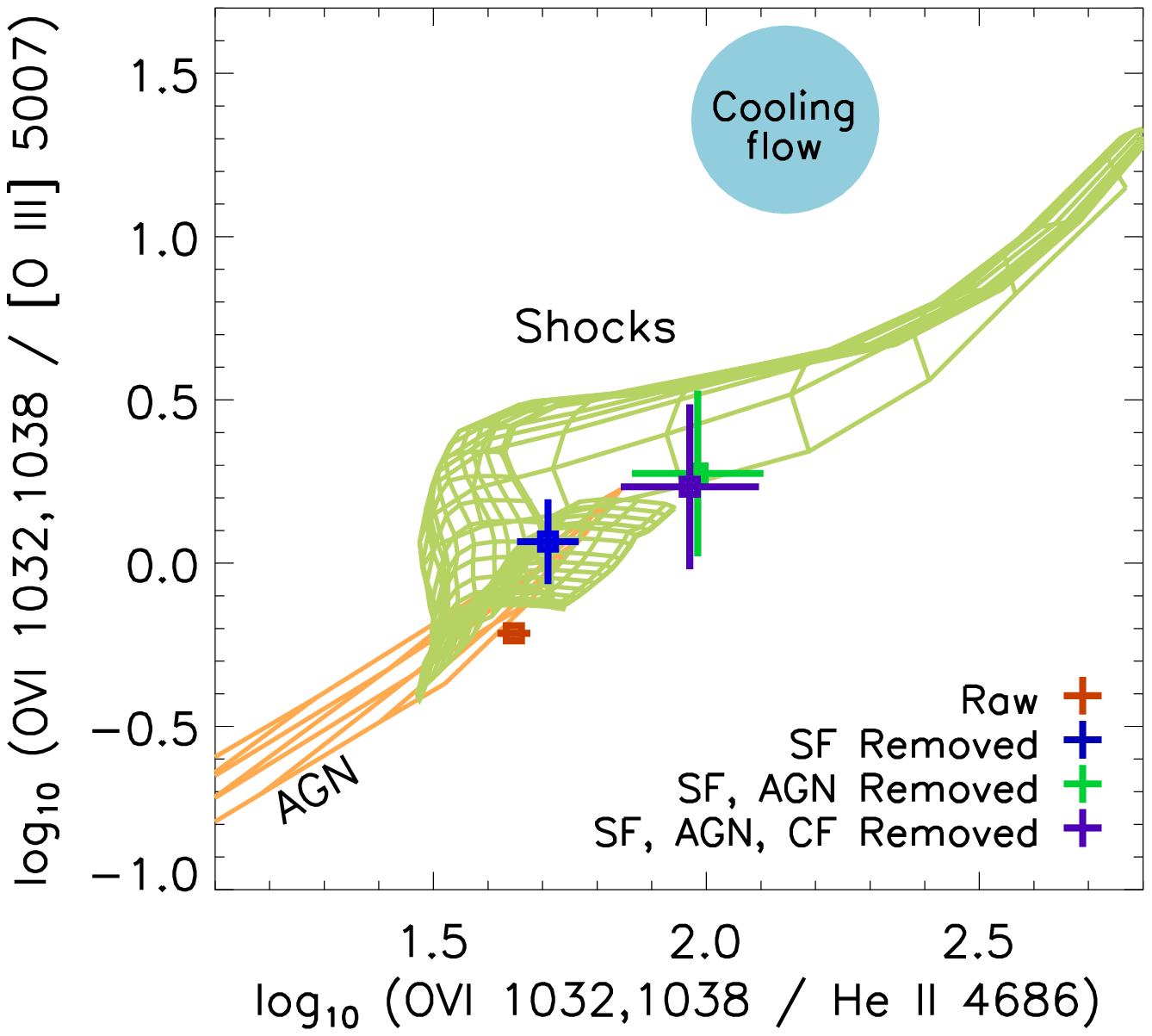}\\
\includegraphics[width=0.48\textwidth]{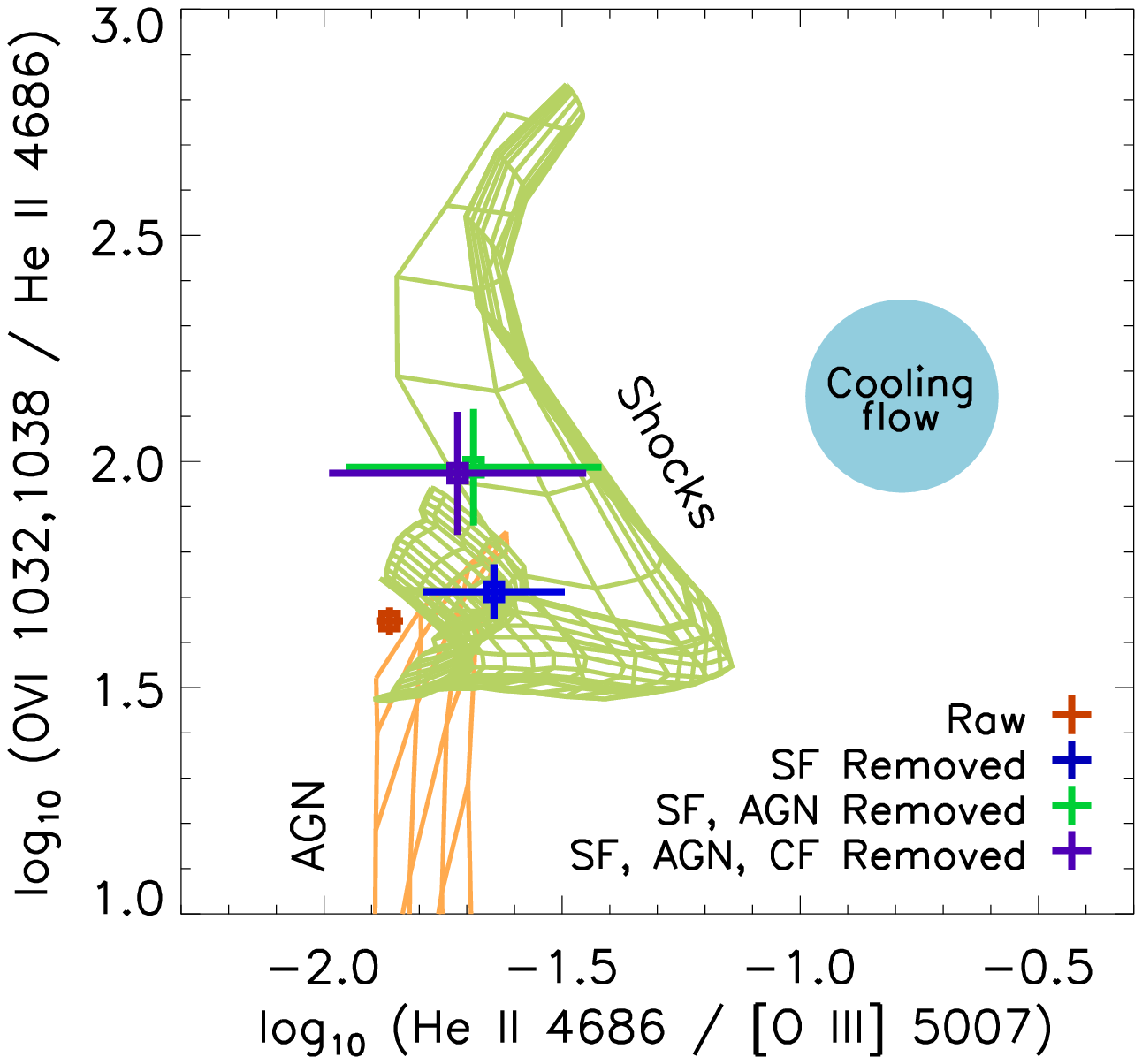}
\end{tabular}
\caption{High-ionization UV-optical emission line ratios for the combined HST-COS apertures shown in Figure \ref{fig:cos_regions}. Overplotted are model expectations for a pure cooling flow \citep{ferland98,chatzikos15}, photoionization from a dust-free AGN \citep{groves04}, and radiative shocks \citep{allen08}. In red we show the measured line ratios from the combined HST-COS and Gemini GMOS spectra shown in Figure \ref{fig:fullspec}. The blue and green points show the residual when photoionization from young stars and a central AGN have been removed, respectively. The increasingly-large errorbars represent our uncertainty in the details of these photoionization models. In purple we show the expected shift if a 5,000 M$_{\odot}$ yr$^{-1}$ cooling flow were subtracted from these data, demonstrating the inability of these data to adequately constrain the cooling rate of the ICM.}
\label{fig:ovi}
\end{figure}

The emissivity curve of O\,\textsc{vi}\,$\lambda\lambda$1032,1038 peaks at $\sim$10$^{5.5}$\,K, making it an excellent probe of gas at intermediate temperatures between the ``hot'' ($>$10$^7$\,K, probed by X-ray emission lines) and ``warm'' ($\sim$10$^4$\,K, probed by optical emission lines) phases. Emission from O\,\textsc{vi} has been detected in several nearby galaxy clusters \citep{oegerle01,bregman06,mcdonald14b}, allowing an independent estimate of the ICM cooling rate. Here, we attempt a similar analysis to these earlier works on the central $\sim$15\,kpc of the Phoenix cluster (see Figure \ref{fig:cos_regions}).

In Figure \ref{fig:sb99model} we show the red-side UV spectra for our two HST-COS pointings. These spectra are modeled using the best-fitting young stellar population described in \S2.2 and \S3.1, along with emission lines including Ly$\beta$, O\,\textsc{vi}\,$\lambda\lambda$1032,1038, and He\,\textsc{ii}\,$\lambda$1085. While the model is, overall, a good match to the data, there are several features present in the data that are not reflected in the models. In the case of the emission at $\sim$1050\AA\, this may be due to time-varying airglow emission. Other features may be absent in the models due to the fact that they are empirically calibrated and lacking the full population of stars included in synthetic optical spectra \citep[see ][]{leitherer14}. We note that He\,\textsc{ii}\,$\lambda$1085 is detected in the northern aperture only, consistent with previously-published optical IFU data \citep{mcdonald14a} which shows that the He\,\textsc{ii}\,$\lambda$4686 peak is offset from the nucleus, due, most likely, to a highly-ionized wind.

We measure a total flux in the O\,\textsc{vi}\,$\lambda\lambda$1032,1038 doublet of $f_{OVI} = 2.46 \pm 0.09 \times 10^{-14}$ erg s$^{-1}$ cm$^{-2}$ and $f_{OVI} = 2.56 \pm 0.10 \times 10^{-14}$ erg s$^{-1}$ cm$^{-2}$ for the northern and southern apertures, respectively. The placement of these apertures was chosen so that the spectra could be added (i.e., the combined throughput never sums to $>$1; see Figure \ref{fig:cos_regions}), meaning that the total O\,\textsc{vi} luminosity in the central $\sim$15\,kpc of the Phoenix cluster is L$_{OVI} = 7.55 \pm 0.20 \times 10^{43}$ erg s$^{-1}$. For comparison, \cite{bregman06} found L$_{\textrm{O}\,\textsc{vi}} \sim 6\times10^{40}$ erg s$^{-1}$ for the Perseus cluster, using UV spectroscopy from the FUSE satellite.
 
Under the naive assumption that 100\% of the O\,\textsc{vi}\,$\lambda\lambda$1032,1038 emission is a result of gas cooling through $\sim$10$^{5.5}$\,K, we can estimate the ICM cooling rate at intermediate temperatures. Using the latest \textsc{cloudy} code \citep{ferland98, chatzikos15}, and assuming initial plasma properties matched to the measured values in the inner $\sim$150 kpc of the cluster ($Z=0.6Z_{\odot}$, kT = 10 keV), we find 
\.{M}$_{\textrm{O}\textsc{vi}} = (\textrm{L}_{\textrm{O}\textsc{vi}} / 1.37\times10^{39}$ erg s$^{-1}$). For the measured luminosity quoted above, this corresponds to a cooling rate of 55,000 M$_{\odot}$ yr$^{-1}$. However, there are additional sources of ionization that may be dominating the line flux here, specifically photoionization from the central AGN and heating from shocks. In \cite{mcdonald14a} we show that both of these ionization sources are contributing to the line flux in high-ionization lines such as [O\,\textsc{iii}]\,$\lambda$5007 and He\,\textsc{ii}\,$\lambda$4863.

In Figure \ref{fig:ovi} we show the combined (both apertures) aperture-matched (between HST-COS and GMOS) line ratios for various sets of high-ionization UV and optical emission lines. The observed, extinction-corrected line ratios are consistent with having origins in either dust-free AGN \citep{groves04} or fast radiative shocks \citep{allen08}. We subtract the contribution to these emission lines from star-forming regions, converting the UV-derived star formation rates to emission line fluxes for a range of stellar photoionization models \citep{kewley01}. The larger uncertainties on the SF-corrected line ratios reflects our uncertainty in the ionization field of these young stars. These revised line ratios, with stellar photoionization contributions removed (Figure \ref{fig:ovi}, blue points), remain consistent with an AGN or shock-heated origin.

\begin{figure*}[htb]
\centering
\includegraphics[width=0.9\textwidth]{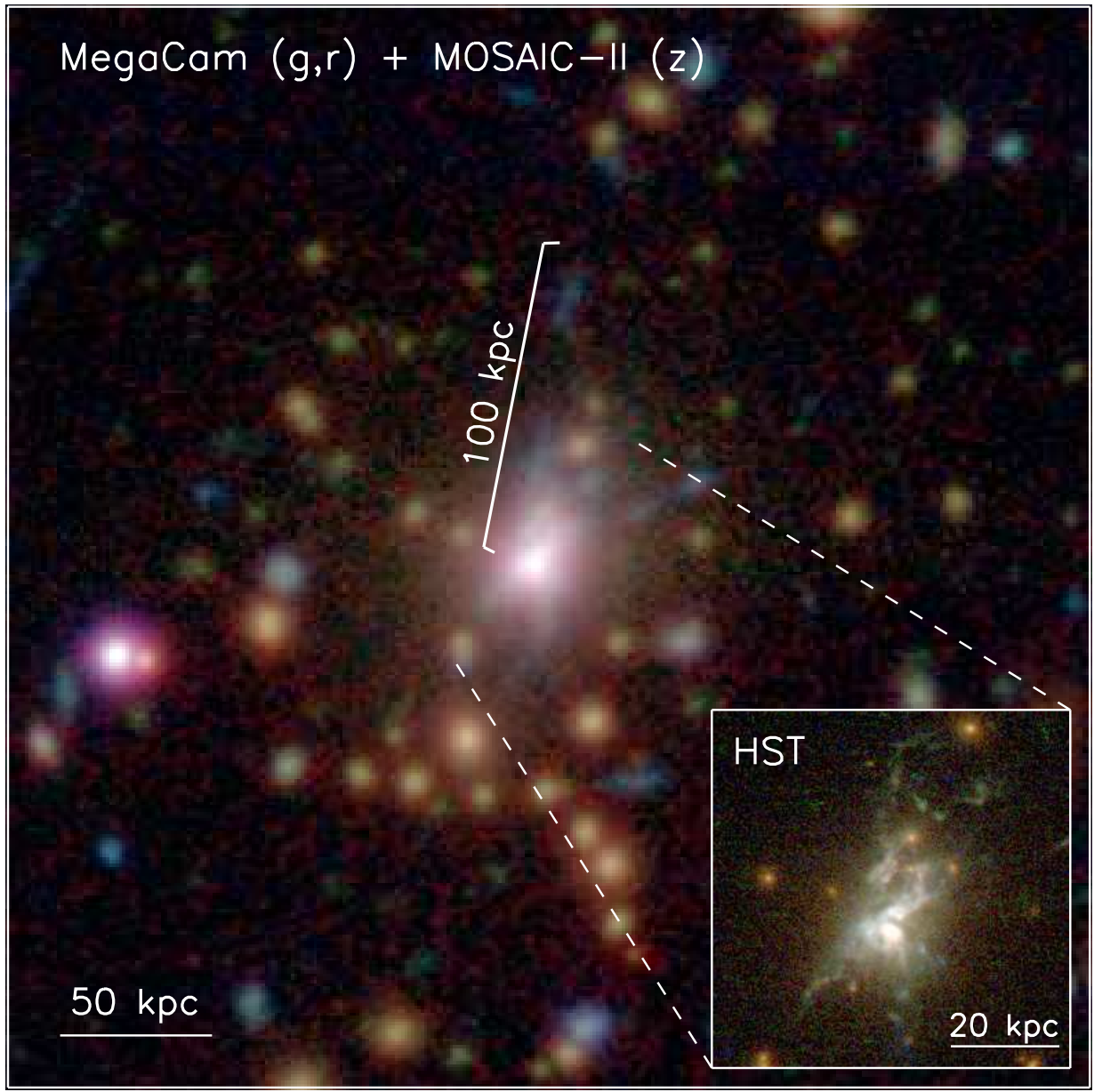}
\caption{Optical $g,r,z$ image of the inner region of the Phoenix cluster (SPT-CLJ2344-4243). The central cluster galaxy is located at the center of the image. Extending radially from this galaxy are several blue filaments, likely sites of ongoing star formation. We highlight the longest filament to the north, which is extended for $\sim$100\,kpc. In the lower right we show a zoom-in of the central galaxy, based on higher-resolution HST data in the same bandpasses. These data were originally presented in \cite{mcdonald13a} and reveal a complex, filamentary morphology in this starburst galaxy.}
\label{fig:megacam}
\end{figure*}

We remove the AGN contribution to the emission lines by extracting the optical spectrum from our GMOS IFU data in a 1$^{\prime\prime}$ wide aperture centered on the nucleus. Assuming a range of AGN photoionization models \citep{groves04} to estimate the O\,\textsc{vi} emission, based on the observed He\,\textsc{ii}$\lambda$4686 flux, we are able to infer the amount of, for example, O\,\textsc{vi} emission coming from the nucleus, without having any actual spatial information from the COS spectroscopy. The inferred contribution to each line from the AGN was subtracted, allowing us to estimate AGN-free line ratios in the optical and UV.. The resulting line ratios, which have both stellar and AGN photoionization removed (Figure \ref{fig:ovi}, green points), remain consistent with radiative shocks with $v \sim 300$ km s$^{-1}$ \citep{allen08}. The large error bars, resulting from our combined uncertainty in the stellar and AGN photoionization fields, are consistent with the full range of magnetic fields tested by \cite{allen08}.  This is consistent with our earlier work \citep{mcdonald14a}, in which we argued that the warm (10$^4$K) gas is predominantly heated by radiative shocks, based on multiple optical line ratio diagnostics (e.g., [O\,\textsc{iii}]/H$\beta$, [O\,\textsc{ii}]/[O\,\textsc{iii}], as well as the gas kinematics. In this earlier work, we showed evidence for a high-velocity, highly-ionized plume of gas extending north from the central AGN, along the same direction as our two COS pointings. Such highly-ionized, high-velocity signatures are typically not observed in low-$z$ clusters, where the warm gas tends to be only weakly ionized.

Figure \ref{fig:ovi} demonstrates that, even after removing ionization contributions from young stars and AGN, the observed high-ionization line ratios are inconsistent with a pure cooling flow model from \textsc{cloudy} \citep{ferland98, chatzikos15}. However, we demonstrate that including a 5000 M$_{\odot}$ yr$^{-1}$ cooling flow to the model (representing only $\sim$10\% of the observed O\,\textsc{vi} flux, purple point in Figure \ref{fig:ovi}) does not significantly change the observed line ratios, which remain consistent with shocks. 
Further, we expect additional ionization from both particle heating \citep[e.g.,][]{ferland09} and mixing \citep[e.g.,][]{fabian11}. 
The former likely contributes at a low level throughout the cluster core. Given that the radio flux of the AGN is fairly typical of BCGs with significantly less star formation, we expect particle heating to be negligible compared to other ionization sources in the extreme environment of the Phoenix cluster. 
On the other hand, mixing of hot and cold gas is likely contributing significantly given the abundance of multiphase gas in this system. Mixing of the hot and cold gas ought to result in an intermediate-temperature phase, which would likely be near $\sim$10$^6$\,K \citep{fabian11}. This would lead to substantial O\,\textsc{vi} emission for relatively low rates of net cooling from the hot to cold phase. 
Thus, we conclude that, as a result of the myriad of additional high-ionization sources, including shocks \citep[e.g.,][]{mcdonald12a}, mixing \citep[e.g.,][]{fabian11}, and particle heating \citep[e.g.,][]{ferland09}, we are unable to constrain the properties of the cooling ICM, which is contributing negligibly to the total O\,\textsc{vi} flux. Additional high-ionization lines such as O\,\textsc{vii} or Fe\,\textsc{xiv} would improve these constraints, perhaps allowing an estimate of the amount of gas cooling through $\sim$10$^6$\,K.

\subsection{Star-Forming Filaments on 100\,kpc Scales}

In \cite{mcdonald13a} we presented high-angular-resolution broadband imaging of the cluster core, revealing complex filaments of star formation on scales of $\sim$40\,kpc. While providing unmatched angular resolution, these data were relatively shallow ($\sim$12m), making the detection of more extended, diffuse emission at large radii challenging. In contrast, the MegaCam data presented, while having inferior angular resolution to the HST data, are considerably deeper, providing substantially-improved sensitivity to faint, extended sources. 

In Figure \ref{fig:megacam} we show a three-color image which combines two MegaCam bands ($g$, $r$) and a redder band ($z$) from the MOSAIC-II camera on the Blanco 4-m telescope. In the cluster core, the new data show that the star-forming filaments initially identified with HST \citep{mcdonald13a} extend significantly further than previously thought. To the south, a pair of filaments extend for $\sim$40 \,kpc each, while to the northwest a third filament extends for $\sim$65\,kpc. The most extended filament extends due north from the central galaxy for $\sim$100\,kpc (highlighted in Figure \ref{fig:megacam}). This is the most extended star-forming filament yet detected in a cool core cluster, exceeding the well-studied filaments in the nearby Perseus \citep[60\,kpc;][]{conselice01,canning14}, Abell~1795 \citep[50\,kpc;][]{cowie83,mcdonald09}, and  RXJ1532.9+3021 \citep[50\,kpc;][]{hlavacek13} clusters. These four filaments, particularly the northern pair, are exceptionally straight, similar to the northern filament in the Perseus cluster. This morphology has been used to argue for relatively low turbulence in the core of the Perseus cluster\citep[e.g.,][]{fabian08}, although counter arguments can be made on the basis of ICM density fluctuations \citep[e.g.,][]{zhuravleva15}.

The filamentary emission is observed in both the $g$ and $r$ bands for all four filaments, providing preliminary evidence that this is continuum, rather than line, emission. At the rest frame of the Phoenix cluster, the $g$-band spans 2400--3300\AA. This wavelength range is relatively free of strong emission lines, with Mg\,\textsc{ii}\,$\lambda$2798 being the only line that may be contributing significantly to the flux in this band. This line should be faint in most scenarios, with the exception of fast ($v \gtrsim 500$ km s$^{-1}$) radiative shocks in a dense ($n\gtrsim100$ cm$^{-3}$) medium.

The total amount of rest-frame UV emission in the outer filaments -- those that were not included in \cite{mcdonald13a} -- is relatively small. For example, the outer 50\% of the northern filament (50--100\,kpc) contributes $\sim$0.5\% of the total blue emission, while the thin northwestern filament contributes an additional $\sim$1\%. Thus, the global estimate of the star formation  is not significantly altered by including these deeper data.

\subsection{X-ray Surface Brightness Maps}

\begin{figure}[htb]
 \centering
\includegraphics[width=0.4\textwidth]{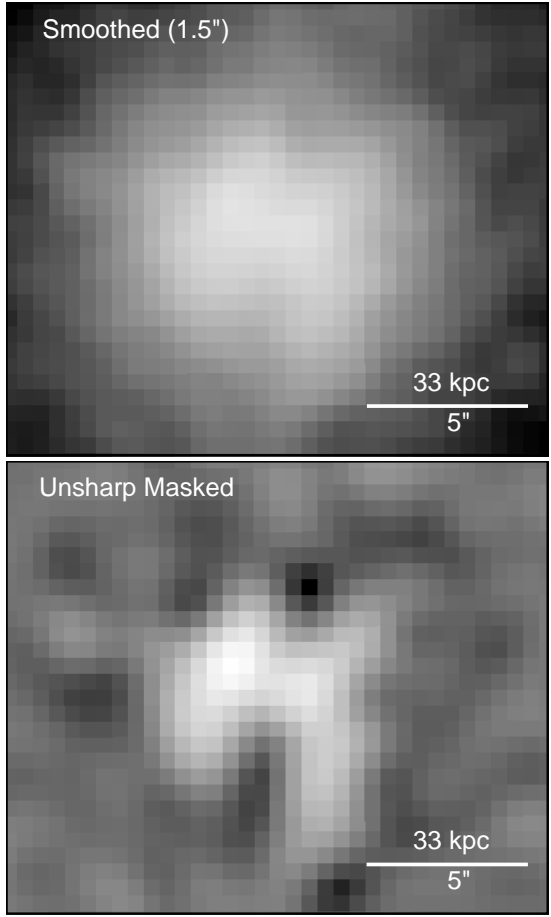}
\caption{Upper panel: Gaussian smoothed (FWHM = 2$^{\prime\prime}$) 0.5--2.0 keV image of the central $\sim$100\,kpc of the Phoenix cluster. This image clearly shows the pair of cavities $\sim$10\,kpc to the north and south of the X-ray peak, despite the lack of any additional processing. Lower panel: Unsharp masked image of the same region as above. This residual image highlights the small-scale structure in the inner region of the cluster, showing a pair of highly-significant cavities to the north and south and overdense regions to the east and west.
}
\label{fig:unsharp}
\end{figure}

\begin{figure*}[htb]
 \centering
\includegraphics[width=\textwidth]{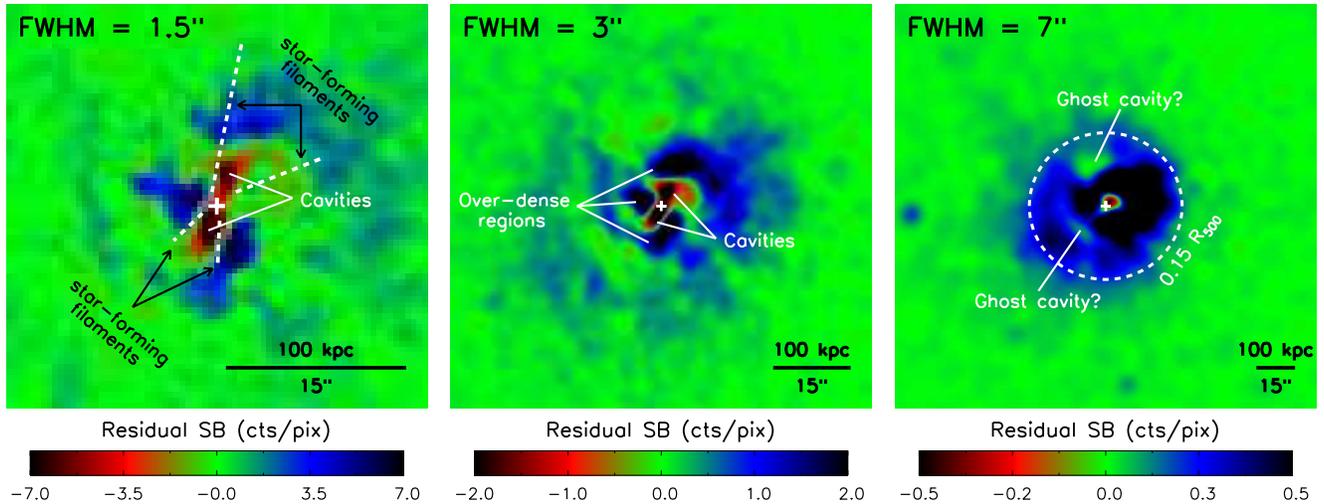}
\caption{Residual X-ray surface brightness images (0.5--2.0 keV) after a double beta model (shared center, ellipticity, position angle) has been subtracted. 
Left: Residual image smoothed with a fine kernel (Gaussian FWHM = 2.2$^{\prime\prime}$). This image highlights the negative residuals (red) near the central AGN (white cross) which are assumed to be X-ray cavities. The dashed radial lines depict the locations of the most extended star-forming filaments (Figure \ref{fig:megacam}). This panel demonstrates that the most extended star-forming filaments appear to avoid the central X-ray cavities.  There are also three regions of enhanced (blue) surface brightness, surrounding the X-ray cavities, which may be sites of enhanced cooling.
Middle: Residual image smoothed with a medium kernel (Gaussian FWHM = 3.5$^{\prime\prime}$). This image is zoomed out by a factor of 3.2 to highlight large-scale features, such as the spiral-shaped overdensities (blue) and the asymmetric central cavities (red). 
Right: Residual image smoothed with a coarse kernel (Gaussian FWHM =  7$^{\prime\prime}$). This image is zoomed out by an additional factor of 2, to show the relative lack of structure on scales larger than $\sim$200\,kpc (0.15R$_{500}$). We highlight a potential set of ``ghost cavities'' \citep[e.g.,][]{mcnamara01} to the north and south, $\sim$100\,kpc from the cluster center. These may be the remnants of a prior epoch of strong radio-mode feedback, although they are only marginally detected (see Table \ref{table:cavities}).
}
\label{fig:residual}
\end{figure*}

In Figure \ref{fig:unsharp} we show a smoothed 0.5--2.0 keV image of the central $\sim$100\,kpc in the Phoenix cluster. Without any additional processing, the pair of cavities in the inner $\sim$10\,kpc, reported initially by \cite{hlavacek14}, are evident. An unsharp masked image of the core (lower panel of Figure \ref{fig:unsharp}) reveals significant structure in the inner $\sim$30\,kpc, with overdense regions surrounding the pair of cavities. The cavities, located $\sim$10\,kpc to the north and south of the central AGN, are detected at a significance of $\sim$25$\sigma$ in this residual image. The unsharp mask technique is only sensitive to substructure on a single angular scale, so it is unsurprising that we detect no large-scale asymmetries beyond the core region in this residual image.

In order to look for ICM substructure on multiple angular scales, we model the 0.5--2.0 keV surface brightness map with a sum of two beta models and a constant background, using \textsc{sherpa}\footnote{\url{http://cxc.harvard.edu/sherpa4.4/}}. The two beta models, meant to represent the core and outer ICM, share a common center, ellipticity, and position angle. We note that, since the X-ray spectrum of the central AGN is highly obscured \citep{mcdonald12c,ueda13}, there is no central point source detected in this soft energy band. During the fit, the cavities (Figure \ref{fig:unsharp}) and point sources were masked to prevent any spurious positive or negative residuals. The best-fitting model was subtracted, with the residual images shown in Figure \ref{fig:residual}. This residual image shows a significant amount of structure in the central $\sim$150\,kpc.
On scales larger than $\sim$200\,kpc (0.15R$_{500}$) there is little structure in the residual image (see rightmost panel in Figure \ref{fig:residual}). 
If the cluster had recently experienced a recent minor merger, one would expect to find large-scale spiral-shaped surface brightness excesses that would be visible on $>$100\,kpc scales in the residual image \citep[e.g.,][]{roediger11, paterno13}. For comparison, the large-scale spiral feature in Abell~2029 \citep{paterno13} represents an excess surface brightness of 22\% at a radius of $\sim$150 kpc. At similar radii in the Phoenix cluster, the surface brightness uncertainty in the 0.5--2.0 keV image is $\sim$15\%, meaning that a similar feature to that observed in Abell~2029 would be at the level of the noise. Thus, we are unable to rule out a recent \emph{minor} merger. In general, the cluster appears relaxed -- an observation corroborated by \cite{mantz15}, who find a marked absence of any large-scale ``sloshing'' features in this system based on an independent analysis, labeling it as one of the most relaxed clusters in the known Universe. 

Directly to the northwest and southeast of the X-ray peak there are a pair of cavities, detected with $S/N = 25$, and consistent in location and size to those reported in \cite{hlavacek14}. The northern cavity appears to be ``leaking'' to the west, while the southern cavity is slightly extended to the east. If these cavities are indeed not confined to a simple bubble morphology, then the estimates of the AGN power derived in previous works \citep{hlavacek14} are significantly underestimating the full power output of the AGN. Interestingly, the four most extended, linear filaments of young stars (Figure \ref{fig:megacam}) appear to trace the AGN outflow, with the northern and southern cavities occupying the space between the two filaments in the respective direction. This may be due to the expansion of the cavity compressing the surrounding gas, leading to more rapid cooling along the edges of these bubbles. 

We note that these two X-ray cavities are at a common distance of $\sim$18\,kpc from the X-ray peak, which is relatively small given the massive size of the cool core ($\sim$100--200 kpc). This may provide the first clue towards understanding the extreme level of star formation in the core of the Phoenix cluster. Under the assumption that these bubbles rise buoyantly,  the distance of the cavity from the radio source can be linked to the age of the AGN outburst. This would signal that the outburst of AGN feedback in the core of the Phoenix cluster is relatively recent ($\sim$10--100 Myr), and may not yet have had time to distribute the energy required to offset cooling to the surrounding ICM. We will return to this line of reasoning in \S4.3.

The residual X-ray images (Figure \ref{fig:residual}) also show a significant amount of excess above a smooth component. We note that this excess is not an artifact induced by the sharp negative residuals (cavities), as these have been masked in the 2D fitting process.
The left-most panel of Figure \ref{fig:residual} shows three distinct regions of excess emission (colored blue). These overdensities may be gas that has been ``pushed aside'' by the inflating bubbles, infalling cool clouds, or low-mass groups that are in the midst of merging with the main cluster. Surrounding the northern cavity, the largest of these overdensities extends over $>$100\,kpc (see central panel of Figure \ref{fig:residual}). The morphology of this overdense region is reminiscent of an infalling cloud of cool gas. Alternatively, the fact that both the northern bubble and the northern overdensity share the same curvature suggests that this may simply be the outer rim of a bubble expanding in a dense, cool core. To classify these with more certainty, we require deep enough X-ray data to produce temperature and metallicity maps.

On larger scales, there may be a pair of ``ghost'' cavities to the north and south of the cluster center. These are located at distances of $\sim$100\,kpc from the X-ray peak and are detected most readily in a heavily-smoothed residual image (right panel of Figure \ref{fig:residual}). We will return to a discussion of the significance of these cavities in \S4.3.

\subsection{Radio Jets, Ionized Outflows, and X-ray Cavities}

\begin{figure}[htb]
\centering
\includegraphics[width=0.45\textwidth]{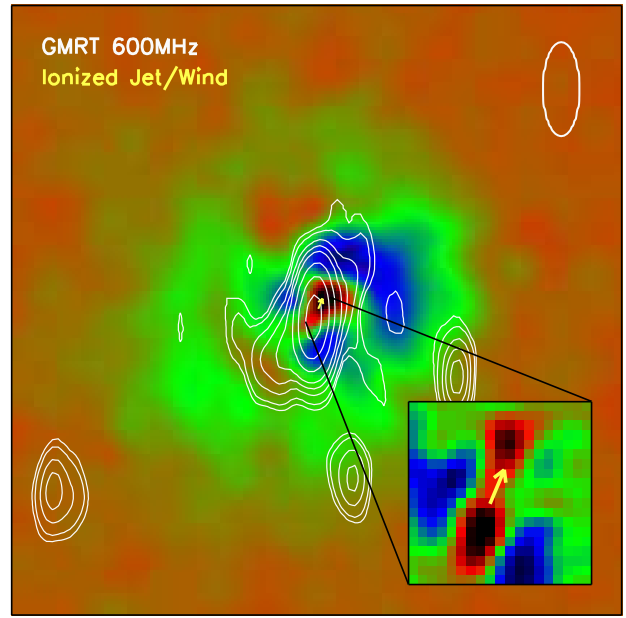}
\caption{This figure shows the smoothed residual maps from Figure \ref{fig:residual} with 600 MHz radio contours overlaid in white. The beam shape of the 600 MHz data is shown in the upper right corner. These contours highlight radio emission centered on the X-ray nucleus and elongated in the north-south direction, in the same direction as the most significant set of cavities. At large radii, the radio emission appears to bend in the direction of the ``leaky'' cavities. The extended radio emission to the southeast appears to coincide with the most extended cavity, at a distance of $\sim$100\,kpc. We also highlight the direction of the ionized wind (inset), discussed in \cite{mcdonald13a}, with a yellow arrow.}
\label{fig:jets}
\end{figure}

In \cite{mcdonald14a}, we present evidence from spatially-resolved optical spectroscopy of a highly-ionized outflow to the north of the central cluster galaxy. This outflow was identified via the presence of a peak in high-ionization optical emission lines (e.g., [O\,\textsc{iii}], He\,\textsc{ii}) roughly 20\,kpc north of the galaxy nucleus. In Figure \ref{fig:jets} (inset), we show that this highly-ionized gas lies in the same direction as the northwestern cavity in the X-ray emission. This may indicate that strong AGN feedback, which is excavating the central cavities, is also heating the multiphase gas, presumably via shocks. This is similar to what is observed in Abell~2052 \citep{blanton11}.

Figure \ref{fig:jets} also shows the extended morphology of the 600\,MHz radio data from GMRT. These data show emission centered on the X-ray peak and central galaxy, extended in the same north-south direction as the cavities. Interestingly, the extended emission spreads to the southeast and northwest, similar to the ``leaking'' cavities shown in Figure \ref{fig:residual}. The most extended emission in the southeast appears to trace the morphology of the potential ``ghost'' cavities, identified in the previous section (see right panel of Figure \ref{fig:residual}). This low-significance cavity may, instead, be an extension of the primary southern cavity, with the radio emission leaking out to the southeast in a similar manner to the northern cavity. This seems to be the case based on the central panel of Figure \ref{fig:residual}. 
Alternatively, the presence of isolated cavities at large radii may signal two distinct bursts of AGN feedback, separated by a relatively small amount of time. We will discuss these scenarios, among others, in \S4. 

The S-shaped morphology of the radio data may also be due to bulk motions of the ICM. \cite{mendygral12} showed that gas motions in the core of clusters could bend MHD jets. Given that the over-dense gas shares a similar morphology, it may be that both the X-ray residuals and the radio morphology can be explained by sloshing of the cool core, as we would expect if this system had undergone a recent minor merger.

We note that, given the large, asymmetric beam in these radio data, it is challenging to say with any certainty how well the radio and X-ray morphologies are related. We await higher angular resolution radio observations to establish the morphological connection between the X-ray cavities and the radio emission.

\subsection{Thermodynamics of the ICM}

\begin{figure}[htb]
\includegraphics[width=0.475\textwidth]{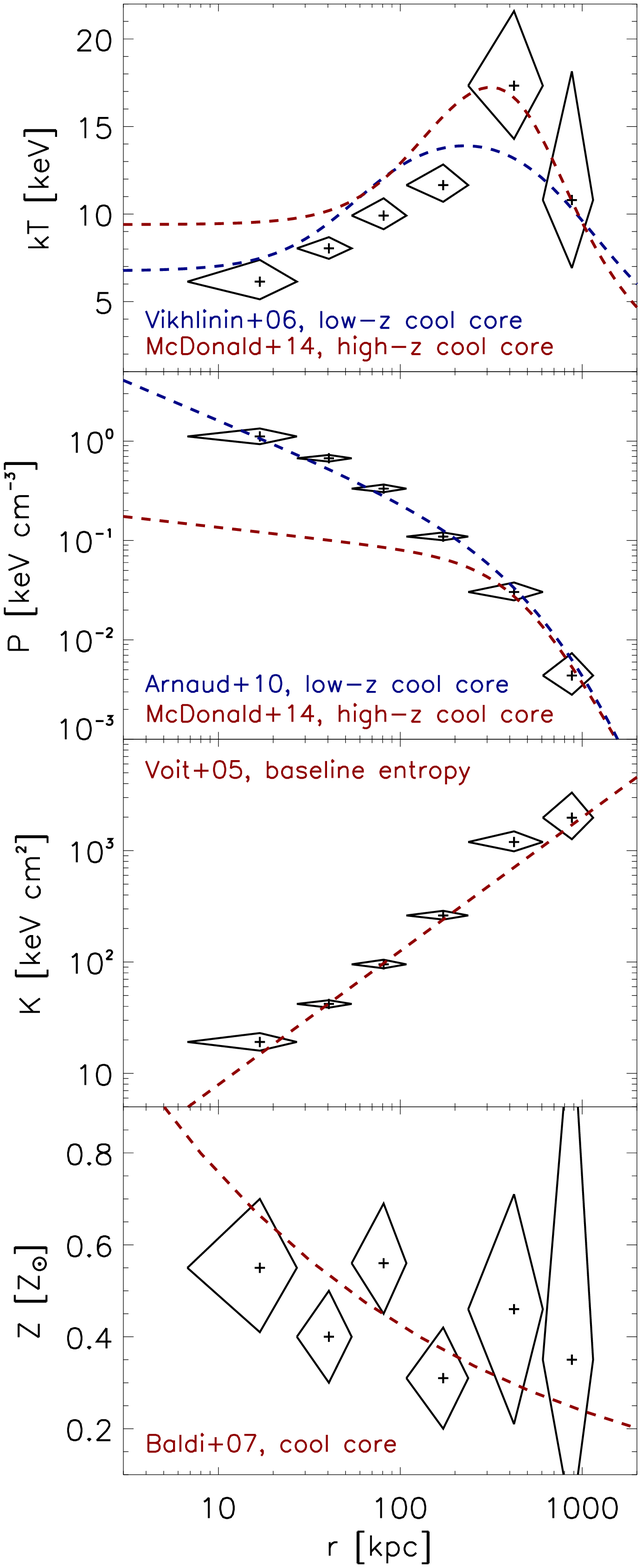}
\caption{Thermodynamic profiles from X-ray spectroscopy. Spectra are deprojected using \textsc{dsdeproj} \citep{sanders07,russell08} and fit with a combination of absorption from neutral gas (\textsc{phabs}) and plasma emission (\textsc{mekal}). Uncertainty regions are shown with diamonds. For comparison, we show expectations (where available) based on averages of low- and high-$z$ clusters in similar mass ranges. The central temperature, pressure, and entropy in Phoenix are at the extreme end for low-$z$ clusters, and are even more extreme compared to other clusters at $z>0.5$ which tend to have less massive cool cores \citep{mcdonald13b}.}
\label{fig:profiles}
\end{figure}

\begin{deluxetable}{c c c c c}[htb]
\tablecaption{Deprojected Thermodynamic Profiles from X-ray Spectra}
\tablehead{
\colhead{Radii [kpc]} &
\colhead{kT [keV]} &
\colhead{P [keV cm$^{-3}$]} & 
\colhead{K [keV cm$^{2}$]} & 
\colhead{Z [Z$_{\odot}$]} 
}
\startdata
   7--  27 &  6.1$^{+ 1.2}_{- 1.0}$ & 1.114$^{+0.225}_{-0.183}$ &     19.2$^{+     3.9}_{-     3.2}$ &  0.55$^{+ 0.15}_{- 0.14}$ \\
  27--  54 &  8.0$^{+ 0.6}_{- 0.6}$ & 0.672$^{+0.054}_{-0.050}$ &     42.1$^{+     3.4}_{-     3.1}$ &  0.40$^{+ 0.10}_{- 0.10}$ \\
  54-- 108 &  9.9$^{+ 1.0}_{- 0.8}$ & 0.332$^{+0.033}_{-0.026}$ &     95.4$^{+     9.5}_{-     7.5}$ &  0.56$^{+ 0.13}_{- 0.11}$ \\
 108-- 236 & 11.6$^{+ 1.2}_{- 0.9}$ & 0.109$^{+0.011}_{-0.009}$ &    262$^{+    26.5}_{-    21.4}$ &  0.31$^{+ 0.11}_{- 0.11}$ \\
 236-- 608 & 17.3$^{+ 4.3}_{- 3.0}$ & 0.030$^{+0.007}_{-0.005}$ &   1200$^{+   294}_{-   210}$ &  0.46$^{+ 0.25}_{- 0.25}$ \\
 608--1148 & 10.8$^{+ 7.3}_{- 3.9}$ & 0.004$^{+0.003}_{-0.002}$ &   1980$^{+  1350}_{-   710}$ &  0.35$^{+ 0.77}_{- 0.35}$ 
 \enddata
\tablecomments{The central 1$^{\prime\prime}$ has been clipped due to the presence of an X-ray point source (central AGN).}
\label{table:spectra}
\end{deluxetable}


In Figure \ref{fig:profiles} and Table \ref{table:spectra} we show the results of our X-ray deprojection analysis (see \S2.2). The measured temperature, entropy, and pressure profiles are roughly consistent with the profiles derived based on a mass-modeling approach in \cite{mcdonald13b}.
In general, the profiles look as expected for a strong cool core, with the temperature dropping by a factor of $\sim$3 between the peak temperature ($r \sim 500$\,kpc, 0.4R$_{500}$) and the central temperature. The temperature profile is similar in shape to the average profiles of low-$z$ \citep{vikhlinin06a, pratt07,baldi07,leccardi08a} and high-$z$ clusters \citep{baldi12,mcdonald14c}, which typically peak at 0.3--0.4R$_{500}$. The factor-of-three drop in the temperature in the inner region is on the high end of what is observed in massive clusters \citep{vikhlinin06a,mcdonald14c}. 

The central pressure (1.1 keV cm$^{-3}$) is exceptionally high -- the highest measured in any cluster to date -- consistent with the expectation for a strong cool core embedded in one of the most massive clusters known. The pressure drops off rapidly with radius, consistent with the expectation from the universal pressure profile \citep[e.g.,][]{arnaud10,planck13,mcdonald14c}. The pressure profile in the Phoenix cluster appears more consistent with $z\sim0$ clusters than other massive clusters at $z\sim0.6$ (see Figure \ref{fig:profiles}).

The deprojected central ($r \lesssim 30$\,kpc) entropy is 19.2 keV cm$^2$, consistent with what is found for nearby cool core clusters \citep[e.g.,][]{cavagnolo09,hudson10}. The entropy profile shows no evidence of excess entropy \citep[e.g.,][]{cavagnolo09} in the inner region, following the ``baseline'' profile of \cite{voit05} from the central bin to the outermost bin at $\sim$1\,Mpc. There is no evidence of flattening in the outer part of the entropy profile, as is seen in many low-$z$ and high-$z$ systems \citep[e.g.,][]{bautz09, walker13, reiprich13, urban14,mcdonald14c}.

The measured metallicity profile is consistent with the declining profile found in low-$z$ clusters \citep[e.g.,][]{degrandi04,baldi07,leccardi08b}. We note that, given the large uncertainties, the metallicity profile is also consistent with a constant value as a function of radius.

The ratio of the cooling time to the freefall time ($t_{cool}/t_{ff}$) in the cores of nearby clusters correlates with the presence of multiphase gas, presumably signaling a link between the warm and hot phases \citep{mccourt12}. Following \cite{gaspari12}, we estimate the ratio of the cooling time to the freefall time in the core of Phoenix using the following equations:

\begin{equation}
t_{cool} = \frac{3}{2} \frac{(n_e + n_i)}{n_en_i\Lambda(T,Z)}
\end{equation}

\begin{equation}
t_{ff} = \left(\frac{2r}{g(r)}\right)^{1/2}
\end{equation}

 \begin{figure}[htb]
 \centering
\includegraphics[width=0.49\textwidth]{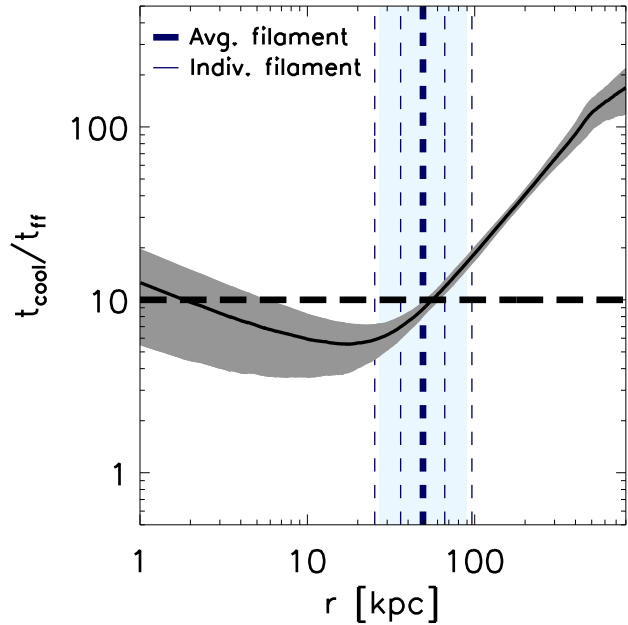}
\caption{Ratio of the cooling time ($t_{cool}$) to the free-fall time ($t_{ff}$) as a function of radius for the Phoenix cluster. The shaded grey band represents our uncertainty, which is primarily driven by uncertainty in the temperature profile. We highlight $t_{cool}/t_{ff} = 10$ with a horizontal dashed line, which appears to be the approximate threshold for condensation in nearby clusters \citep{mccourt12,sharma12,gaspari12}. We also highlight the maximum extent of four distinct star-forming filaments (Figure \ref{fig:megacam}) with thin vertical lines, and the average extent with a thick vertical line. The range of filament radii is shown as a shaded blue region. This figure demonstrates that the condensation of the cooling ICM may be fueling star formation out to radii of $\sim$100\,kpc in this extreme system.}
\label{fig:tctff}
\end{figure}

\noindent{}We estimate g(r) assuming that the core is in hydrostatic equilibrium, and the cooling function ($\Lambda(T,Z)$) following \cite{sutherland93}, assuming solar metallicity and $n_i = 0.92n_e$. The resulting $t_{cool}/t_{ff}$ profile is shown in Figure \ref{fig:tctff}. The minimum value of $t_{cool}/t_{ff} \sim 5$ is reached $\sim$20\,kpc from the central AGN, which corresponds to the radius within which the bulk of the star formation is contained \citep[see Figure \ref{fig:megacam} and][]{mcdonald13a}. This minimum value is on par with the minimum value of $t_{cool}/t_{ff}$ in some nearby cool core clusters, which can vary from $\sim$4 to $\sim$20 \citep{voit14b}. However, such low values of $t_{cool}/t_{ff}$ are typically reached at $r<10$\,kpc \citep{gaspari12}.

Interestingly, $t_{cool}/t_{ff}$ remains below 10  (the approximate threshold for ICM condensation in nearby clusters \citep{mccourt12,sharma12,gaspari12}) out to a radius of $\sim$60\,kpc, which is nearly the maximum extent of the star-forming filaments shown in Figure \ref{fig:megacam}. This, combined with the presence of large-scale star-forming filaments, suggests that condensation of the cooling ICM is important in this system out to exceptionally large radii.

We will return to this discussion of the cluster's thermodynamic properties later, to quantitatively compare cooling and feedback processes in the cluster core.

\section{Discussion}
We have presented new X-ray, ultraviolet, optical, and radio data on the core of the Phoenix cluster (SPT-CLJ2344-4243), providing the most complete picture of this extreme system to date. Below, we discuss the implications of these new data in the context of previous observations and interpretations \citep{mcdonald12c,mcdonald13a,ueda13, mcdonald14a, hlavacek14}.

\subsection{A Revised Estimate of the Central Galaxy Star Formation Rate}
In \cite{mcdonald13a}, we calculate the star formation rate (SFR) in the core of the Phoenix cluster based on rest-frame near-UV imaging from HST, obtaining a value of 800 M$_{\odot}$ yr$^{-1}$. In this earlier work, we assume that, in the absence of dust, the slope of the UV spectral energy distribution (SED) should be flat \citep{kennicutt98}. This allowed us to compute an extinction correction based on the UV color. However, the amount of UV extinction and the age of the stellar population are degenerate with regards to the slope of the UV SED. Ideally, we would like an alternate, age-free, estimate of the reddening, which we obtained via Balmer line ratios in \cite{mcdonald14a}. This reddening map showed a peak value of $E(B-V) = 0.5$ at the central AGN, dropping to $E(B-V) \sim 0.1$ in the most extended filaments. Applying this reddening map to the rest-frame near-UV ($\sim$3000\AA) imaging from HST, assuming an extinction curve from \cite{cardelli89} with $R_V = 3.1$ and a relation between the UV luminosity and SFR from \cite{kennicutt98}, yields a revised SFR of 690 M$_{\odot}$ yr$^{-1}$.

This estimate of the star formation rate may be artificially high. \cite{calzetti94} show that starburst galaxies tend to lack the rapid rise in the UV extinction curve that is characteristic of the Milky Way and LMC/SMC extinction curves. This is confirmed in Figure \ref{fig:fullspec}, where we show that a shallow, featureless extinction curve \citep[e.g.,][]{calzetti94} provides a much better match to the UV-through-optical spectrum of the central galaxy in the Phoenix cluster. Using the best-fit instantaneous-burst model from Figure \ref{fig:fullspec}, and averaging over the lifetime of the burst (4.5 Myr), we infer a reduced star formation rate of 490 M$_{\odot}$ yr$^{-1}$. 

As we show in Figure \ref{fig:cos_regions}, the throughput of the combined HST-COS apertures is only $\sim$100\% in the inner region of the starburst. Thus, an accurate estimate of the total SFR based on these data requires an aperture correction. Based on rest-frame near-UV HST imaging from \cite{mcdonald13a}, we find that only $\sim$26\% of the UV continuum emission is missed by the combined HST-COS apertures. Assuming that this missing flux originates from regions with similar extinction, we would infer an aperture-corrected star formation rate of 660 M$_{\odot}$ yr$^{-1}$. More likely, the extinction in the extended filaments is lower than in the inner few kpc. Assuming a much more conservative extinction in the aperture correction ($E(B-V) = 0.15$, representing the minimum value measured in \cite{mcdonald14a}) yields SFR~=~570 M$_{\odot}$ yr$^{-1}$. Assuming that the correct answer lies somewhere between these extrema, we infer that the aperture- and extinction-corrected, time-averaged star formation rate for the central galaxy in the Phoenix cluster is $610 \pm 50$ M$_{\odot}$ yr$^{-1}$, where the uncertainty here represents only the uncertainty in the extinction and aperture corrections. We note that this UV-based estimate is consistent with the recent IR-based estimate from \cite{tozzi15} of $530 \pm 50$ M$_{\odot}$ yr$^{-1}$, suggesting that these different methods may be converging on the right answer.

The dominant systematic uncertainty in this estimate is how, exactly, the star formation has proceeded. For example, an instantaneous burst of average age 2 Myr, which is preferred by the UV-only spectral fit (Figure \ref{fig:ovi}), would have a time-averaged star formation rate of 1200 M$_{\odot}$ yr$^{-1}$ \emph{before aperture correction}. On the other hand, if the star formation has been proceeding at a constant level for the past 15 Myr, the time-averaged rate can be as low as $\sim$300 M$_{\odot}$ yr$^{-1}$ (Figure \ref{fig:fullspec}). This uncertainty could be reduced by obtaining deep, rest-frame near-UV ($\sim$2000\AA) spectroscopy. At these wavelengths, the stellar population synthesis models are considerably more mature and well-tested, which should lead to overall better fits to the data and a tightening of the allowed parameter space.

The Megacam data presented here, while significantly deeper, do not provide additional constraints on the global star formation rate, since the bulk of the UV emission is contained in the inner $\sim$30\,kpc. Instead, these data provide improved constraints on the extent and large-scale morphology of the star-forming filaments. 


\subsection{The Origin of the Star-Forming Filaments}

In the central $\sim$30\,kpc, the UV emission is highly asymmetric (Figure \ref{fig:megacam}), indicative of a vigorous, turbulent starburst. However, at large radii the young stars are oriented along thin, linear filaments, akin to systems like Perseus \citep{conselice01} and Abell~1795 \citep{mcdonald09}. In \cite{mcdonald12c} we demonstrate that the fuel for the observed star formation must originate within the cluster core. A scenario in which cold gas is brought into the core via infalling gas-rich galaxies and/or groups is unfeasible for this system, given the extreme ICM density (efficient ram-pressure stripping) and amount of gas needed to fuel such a starburst (i.e., multiple gas-rich compact groups). 
A popular scenario for the origin of star-forming filaments in nearby cool core clusters involves cool gas being drawn from the cluster core in the wake of buoyant radio-blown bubbles \citep[e.g.,][]{fabian03b,churazov13}. Figures \ref{fig:megacam} and \ref{fig:residual} demonstrate that the most extended filaments surround, rather than trail, the most significant set of cavities, inconsistent with the uplift scenario. Indeed, the fact that the star-forming filaments extend beyond these cavities in radius suggests that the cool gas was present before the current epoch of feedback. However, this does not rule out a scenario in which these filaments were uplifted by a previous episode of feedback, with the bubbles responsible for this action having risen to such radii that they are undetectable.

It may be that no additional mechanism is necessary to produce the extended, star-forming filaments observed in the core of the Phoenix cluster. 
Recent work by \cite{mccourt12}, \cite{sharma12}, and \cite{gaspari12} has shown that local thermodynamic instabilities can develop when the ratio of the cooling time to the free-fall time ($t_{cool}/t_{ff}$) is less than 10. 
In Figure \ref{fig:tctff}, we show this ratio as a function of radius for the Phoenix cluster, assuming hydrostatic equilibrium in the computation of $t_{ff}$. 
The cooling time is shorter than 10 times the free-fall time over an unprecedented $\sim$60 kpc in radius, suggesting that the star-forming filaments may be fueled by local thermodynamic instabilities in the hot ICM, which rapidly condense and then ``rain'' down onto the central cluster galaxy \citep{voit15}.
Further, we expect a map of $t_{cool}/t_{ff}$ to be asymmetric, since the gravitational potential is roughly spherically symmetric (in the core), while the gas density is not (see Figure \ref{fig:residual}). For example, the spiral-shaped overdensity at $\sim$50\,kpc north of the cluster center will likely have $t_{cool}/t_{ff} < 10$, since the azimuthally-averaged value at that radius is $t_{cool}/t_{ff}\sim10$ and $t_{cool} \propto 1/n_e$. Thus, despite the fact that $t_{cool}/t_{ff} > 10$ at $r>60$\,kpc, there are likely overdense regions in the ICM with $t_{cool}/t_{ff} < 10$ out to the full extent of the star-forming filaments at $r\sim100$\,kpc.

If the star-forming filaments did indeed condense out of the hot ICM, there ought to be evidence of cooling in the X-ray spectrum. 
Using the latest \emph{Chandra} data, we estimate the classical cooling rate following \cite{white97}, using the equation:

\begin{equation}
\dot{M}(i) = \frac{L_X(i)}{h(i) + \Delta\phi(i)}  - \frac{[\Delta\phi(i) + \Delta h(i)]\sum_{i^{\prime}=1}^{i^{\prime}=i-1}\dot{M}(i^{\prime})}{h(i) + \Delta\phi(i)}
\end{equation}

where the i and i$^{\prime}$ indices refer to given annuli, $L_X(i)$ is the bolometric X-ray luminosity in a given annulus, $h(i) \equiv \frac{5}{2}kT(i)/\mu m_p$ is the energy per particle-mass of the hot gas for a given annulus, and $\Delta\phi(i)$ is the change in gravitational potential across shell $i$. The first term on the right corresponds to the cooling rate in the absence of any additional heating, while the second term on the right is the correction factor to account for the fact that infalling gas will simultaneously be heated gravitationally. Carrying out this sum in the inner 100\,kpc, we estimate a classical (luminosity-based) cooling rate of 3300 $\pm$ 200 M$_{\odot}$ yr$^{-1}$, consistent with the estimate of 2700 $\pm$ 700 M$_{\odot}$ yr$^{-1}$ from \cite{mcdonald13a}. 
%
Assuming the star formation rate of 613 M$_{\odot}$ yr$^{-1}$ from \S4.1, based on the most realistic reddening and stellar population model, this implies that $\sim$20\% of the predicted cooling flow is converted into stars. This number is very close to the expected star formation efficiency from cool gas \citep[10--15\%;][]{mcdonald11b,mcdonald14b}, leaving open the possibility that cooling at the hot phase may be $\sim$100\% efficient. 
However, the deprojected X-ray spectra in the inner 50\,kpc show no evidence of a cooling flow, with a single-temperature model providing an excellent fit ($\chi^2$ = 128.75 for 128 degrees of freedom). 
Likewise, recent \emph{XMM-Newton} observations published by \cite{tozzi15} find relatively weak cooling signatures in the spectrum at temperatures of 0.3--3.0 keV, with measurements by the MOS ($620^{+350}_{-240}$ M$_{\odot}$ yr$^{-1}$) and PN ($210^{+145}_{-115}$ M$_{\odot}$ yr$^{-1}$ instruments yielding values significantly lower than the luminosity-based estimate.
This may be because AGN feedback has recently halted cooling at high temperatures (see \S4.3), with the star-forming filaments being fueled by previously-cooled gas. Indeed, in Figure \ref{fig:fullspec}, we show that the optical spectrum in the vicinity of the Balmer break is consistent with a recently-quenched starburst. Alternatively, a significant amount of cooling can be ``hidden'' in the X-ray spectrum, since all of the strong cooling lines at low temperature \citep[O\,\textsc{vii}, O\,\textsc{viii}, Fe\,\textsc{xvii}--\textsc{xx};][]{peterson06} are redshifted to $\lesssim0.6$\,keV where systematic uncertainties in the \emph{Chandra} effective area correction are high due to contamination \citep{odell13}. Further, the disagreement between the spectroscopic cooling rates derived by the pn and MOS detectors on \emph{XMM-Newton} \citep{tozzi15} suggest that systematic uncertainties in our understanding of the soft X-ray response of these detectors (on both \emph{Chandra} and \emph{XMM-Newton}) will dominate any estimate of the spectroscopic cooling rate.
Thus, while there is no direct evidence for cooling in the low-resolution X-ray spectrum, we can not rule out the hypothesis that the star-forming filaments are being fueled by local thermodynamic instabilities in the ICM.

The \emph{Chandra} X-ray data presented here offer further support for a local fuel supply. The short central cooling time ($t_{cool} < 10^8$ yr) falls below the threshold for the onset of cooling instabilities in the hot atmosphere \citep[e.g.,][]{rafferty08}. It also meetings the $t_{cool}/t_{ff}$ criterion for cooling instabilities \citep[e.g.,][]{mccourt12,sharma12,gaspari12,voit15}. However, the level of star formation lies below that expected from unimpeded cooling.  Phoenix harbors a powerful central radio-AGN that is apparently reducing the cooling rate by $\sim$80\%, operating in a self-regulating feedback loop. The absence of spectral signatures of cooling through $10^6$\,K (i.e., Fe\,\textsc{xvii}, O\,\textsc{vii}) as in other clusters \citep{peterson06}, is inconsistent with simple, isobaric cooling models, indicating that the gas may be cooling through other channels (e.g., mixing) or has recently been quenched \citep[e.g.,][]{li14}. Future observations with high-resolution X-ray grating/microcalorimeter spectrometers (e.g., Astro-H) will help to break this degeneracy by providing firm constraints on the rate of radiative cooling from high ($\sim$10$^8$K) through low ($\sim$10$^6$K) temperatures for a sample of nearby, cool core clusters.

\subsection{Mechanical AGN Feedback}

\begin{deluxetable*}{c c c c c c c c c c c}[htb]
\tablecaption{Properties of X-ray Cavities}
\tablehead{
\colhead{Cavity} &
\colhead{r [$^{\prime\prime}$]} &
\colhead{r [kpc]} &
\colhead{a [$^{\prime\prime}$]} &
\colhead{a [kpc]} &
\colhead{b [$^{\prime\prime}$]} &
\colhead{b [kpc]} &
\colhead{pV$^1$ [10$^{59}$ erg]} & 
\colhead{t$_{buoy}$ [10$^7$ yr]} & 
\colhead{P$_{cav}$$^2$ [10$^{45}$ erg s$^{-1}$]} &
\colhead{S/N$^3$}
}
\startdata
\multicolumn{2}{l}{\emph{Central Cavities}}\\
North & 2.6 & 17.3 & 1.7 & 11.3 & 1.3 & 8.6 & 1.9 -- 2.5 &  2.1 -- 5.1 & 0.8 -- 2.5 & 24.7\\
South & 2.6 & 17.3 & 2.2 & 14.6 & 1.3 & 8.6 & 2.5 -- 4.2 & 1.8 -- 5.8 & 0.9 -- 4.7 & 26.2\\
\\
\multicolumn{3}{l}{Total (inner)} & & & & & 4.4 -- 6.7 & & 1.7 -- 7.2 & \\\\
\multicolumn{3}{l}{\emph{Potential Ghost Cavities}}\\
North & 17.0 & 114.6 & 6.6 & 44.8 & 4.6 & 31.1 & 18 -- 26 & 8.7 -- 13 & 1.8 -- 3.8 & 7.4\\
South & 14.4 & 97.2 & 8.6 & 58.4 &  2.9 & 19.3 &  11 -- 34  & 6.4 -- 14 & 1.0 -- 6.7 & 8.1\\
\\
\multicolumn{3}{l}{Total (outer)} & & & & & 29 -- 60 &  & 2.8 -- 10.5 & 
\enddata
\tablecomments{\\$^1$: Range reflects uncertainty in the three-dimensional bubble shape.  \\$^2$: Range reflects combined uncertainty in three-dimensional bubble shape and bubble rise time (assuming buoyant rise).  \\$^3$: Based on residual image (Figure \ref{fig:residual}). Ratio of cavity depth to noise level at the same radius.}
\label{table:cavities}
\end{deluxetable*}

 \begin{figure}[b]
 \centering
\includegraphics[width=0.45\textwidth]{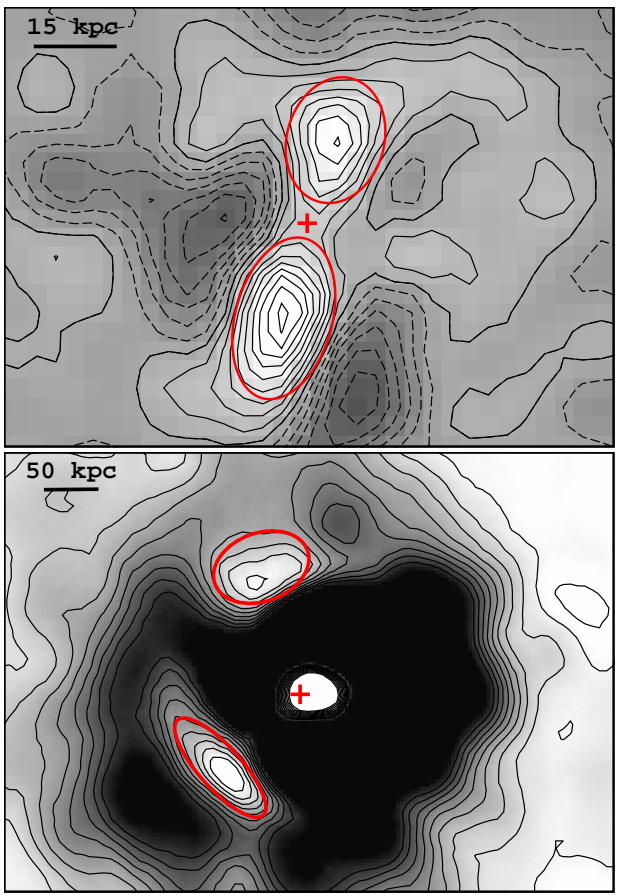}
\caption{Upper panel: Smoothed, residual 0.5-2.0 keV image of the inner $\sim$50~kpc of the Phoenix cluster (see Figure \ref{fig:residual}). Contours highlight structure in the positive (dashed) and negative (solid) residuals. The red cross highlights the position of the central AGN, while red lines show the size and shape of the ellipses used to determine cavity energetics. Lower panel: Similar to upper panel, but zoomed out by a factor of $\sim$5 to show the positions of the extended ``ghost'' cavities. These cavities are also visible in Figure \ref{fig:residual}.}
\label{fig:cavs}
\end{figure}

To estimate the enthalpy released by the radio jets, $E=4PV$, we 
measured the cavity sizes and surrounding pressures \citep[e.g.,][]{churazov00,mcnamara00}. The X-ray cavities are seen in the raw image with well-defined, elliptical shapes.  We measured their locations and sizes using the residual image after smoothing it with a gaussian kernel (FWHM = 3 pixels).  Their radial centroids,  projected major and minor axes, and energetics are presented in Table \ref{table:cavities} and shown in Figure \ref{fig:cavs}.  We assume the cavities are ellipsoidal volumes with axes perpendicular to the plane of the sky and equal to the projected major axes (upper limit) and minor axes (lower limit).   The mean jet power assumes the cavities rose buoyantly
to their current locations in the plane of the sky following \citep{birzan04,mcnamara07}.

For projected distances of ~17 kpc, we find a rise time of $\sim1.6\times10^7$ yr,
assuming the bubbles rise at the sound speed, to a more likely $\sim 2-6 \times 10^7$ yr,
assuming the bubbles rise buoyantly \citep{birzan04}. These figures depend on the local gravitational acceleration measured from the X-ray pressure profile, assuming hydrostatic equilibrium.  The gas 
pressure and mean gas density at the locations of the cavities were found to be $3\times 10 ^{-9}~
\rm erg~cm^{-1}$ and $0.15~\rm cm^{-3}$, respectively. For a total cavity enthalpy of $4pV= 1.8-2.7 \times 10^{60}$ erg, we find a mean jet power of $1.7-7.2 \times 10^{45}~\rm erg~s^{-1}$. This mechanical power places Phoenix among of the most energetic AGN outbursts known. The total radio luminosity (10 MHz -- 10 GHz), assuming a constant spectral index of $\alpha=-1.35$ \citep{mcdonald14a}, is $3.6\times10^{43}$ erg s$^{-1}$, or roughly one percent of the jet power, consistent with other nearby galaxy clusters \citep{birzan08}.

The total mechanical power output measured in the core of the Phoenix cluster is $\sim2-7\times10^{45}$ erg s$^{-1}$. This is a factor of $\sim$4 times less than that quoted by \cite{hlavacek14}. This large difference is, for the most part, a result of shallower X-ray data in the earlier study, which resulted in a factor of $\sim$2 change in pressure and a factor of $\sim$4 change in local gravitational acceleration. The new estimate of the mechanical power lies among the highest measured in a cluster core the core, but is still a factor of $\sim$2--4 times less than the energy required to offset cooling ($L_{cool, r<100kpc} = 9.6\pm0.1 \times 10^{45}$ erg s$^{-1}$). Given the extreme classical cooling rate in the inner 100 kpc (3300 M$_{\odot}$ yr$^{-1}$), even a small mismatch in energetics over a short time could lead to the observed star formation rate of $\sim$600 M$_{\odot}$ yr$^{-1}$ in the central galaxy. 

In Figures \ref{fig:residual} and \ref{fig:cavs} we show the locations of two potential ``ghost'' cavities, at radii of $\sim$100\,kpc. These potential cavities are detected with signal-to-noise of $\sim$7 and $\sim$8 (Table \ref{table:cavities}) and are along the same north-south direction as the radio jets. Due to the lower detection significance, it is possible that only one (or neither) of these cavities are real. However, by computing the total cavity power assuming they are real, we place a rough upper limit on the total amount of power in previous outbursts. Following the same procedures as above, we estimate a total cavity power of $2.8-10.5 \times 10^{45}$ erg s$^{-1}$, or anywhere from a third to 100\% of the cooling luminosity. This range overlaps well with the energy in the ongoing outburst at small radii, suggesting that the mechanical power has been roughly constant over the past two epochs of feedback, spanning $\sim$100 Myr. 

In summary, there is strong evidence for ongoing mechanical-mode feedback in the inner $\sim$20 kpc of the Phoenix cluster, and weaker, but still convincing, evidence for a prior epoch of feedback roughly 100 Myr ago. These episodes of feedback have highly uncertain energetics (factor of a few in P$_{cav}$), but are consistent with the energy required to offset cooling losses. This scenario is further supported by the fact that the cooling time at $r>10$kpc is longer than the apparent duty cycle of this AGN ($\sim$100 Myr), suggesting that perfect balance in a given feedback epoch is not necessary at larger radii. Future studies with significantly deeper X-ray and low-frequency radio data will provide a more clear picture of both the duty cycle and energetics in these radio-mode outbursts.

\subsection{A Transitioning AGN?}

Phoenix is unusual in that its high radio mechanical power is accompanied by a bright
quasar with X-ray photon luminosity $5.6\times 10^{45}~\rm erg~s^{-1}$.  This high X-ray flux is relatively stable, with no significant fluctuations observed between the initial Chandra and Suzaku observations in 2011 and 2012, and the most recent Chandra observations in 2014.
Radio AGN with comparable radiative and mechanical powers are uncommon, as are radiatively-efficient AGN at the centers of clusters \citep{russell10,osullivan12,walker14}.  
The Eddington luminosity for a $10^{9}~\rm M_\odot$ black hole is $1.2\times10^{47}$ erg s$^{-1}$, which would place Phoenix at a few percent of the Eddington rate, close to the expected transition between radio-mechanical AGN and quasars \citep{churazov05,russell13}.  Standard AGN theory posits
that the form of power output depends on the specific accretion rate: objects accreting
above a few percent of the Eddington rate are dominated by radiation and those below
are dominated by mechanical outflows.  That Phoenix exhibits the characteristics of both a quasar and a radio galaxy, suggesting that it may be in transition between the two states.

Based on their discovery of neutral iron emission 
from the AGN using the Suzaku observatory, \cite{ueda13} have argued that the BCG is
a Type 2 quasar with a high absorption column density.  Assuming a bolometric correction factor of 130 \citep{marconi04}, they argue that the total unabsorbed AGN power would be 
$6 \times 10^{47}$ erg s$^{-1}$.   This figure implies a black hole accretion rate
exceeding $100~\rm M_\odot ~\rm yr^{-1}$, which would be the highest rate known in a brightest
cluster galaxy.  Such a high bolometric correction implies a high infrared luminosity, where most of the radiation from the buried AGN should be escaping. The total infrared luminosity emerging from the quasar and surrounding star formation is only $L_{IR} = 3.7 \times 10^{46}$ \citep{mcdonald12c}. This luminosity is comparable to both the mechanical and X-ray energy fluxes, but lies two orders of magnitude below the bolometric luminosity quoted by \cite{ueda13}. We conclude that the bolometric correction to the nuclear X-ray luminosity is at most a factor of a few to ten, consistent with the lower values reported by \cite{vasudevan} for high-luminosity AGN.  Therefore the mechanical power of Phoenix's radio jets is comparable to the radiation emerging from the quasar, suggesting that it may currently be undergoing a transition from ``quasar--mode'' to ``radio--mode''.

\subsection{The Radio Mini-Halo}
The 610 MHz image of the core of the Phoenix cluster previously revealed the presence of a central compact radio source, associated with the BCG, and diffuse $400-500$~kpc emission surrounding this central source \citep{vanweeren14}. The diffuse emission is classified as a radio mini-halo and, at a redshift of  $0.596 \pm 0.002$, is the most distant known. The integrated flux density of  $17\pm5$~mJy for the mini-halo results in a  1.4~GHz radio power of  $(10.4\pm3.5)~\times 10^{24}$ W~Hz$^{-1}$, scaling with a spectral index of $-1.1$.

Considering the lifetimes of the radio emitting electrons and the large extent of the mini-halos, a form of in-situ cosmic ray (CR) production or re-acceleration is required. An explanation for the presence of CR electrons is that turbulence in the cluster core re-accelerates a population of relativistic fossil electrons  \citep{gitti02,gitti04b}. These electrons could, for example, have originated from the central radio source, associated with the BCG.
It has been suggested that gas sloshing, induced by a minor merger event, could generate the required turbulence. This is supported by the observed spatial correlation between the morphologies of some mini-halos and spiral-like sloshing patterns in the X-ray gas of other nearby clusters \citep{mazzotta08}.
Numerical simulations \citep[e.g.,][]{zuhone13} provide further support for this model. 
The X-ray residual maps shown in Figure \ref{fig:residual} do show some spiral structure reminiscent of sloshing excited by a minor merger, suggesting that Phoenix cluster may have undergone a minor merger recently. The lack of any large-scale asymmetry (rightmost panel of Figure \ref{fig:residual}) implies that any merging system must be relatively small compared to the main cluster. 

Due to the mass of this system, the necessary turbulent velocities to accelerate CRs can be achieved by the relatively-smooth accretion of low-mass systems, rather than a single, recent event. Since the Phoenix cluster is amongst the most massive clusters known \citep[$M_{500} = 1.3\times10^{15}$ M$_{\odot}$;][]{mcdonald12c}, and turbulent velocities scale with halo mass (based on structure formation theory), it is feasible that the Phoenix cluster maintains a high enough turbulent velocity in its core to continuously re-accelerate a population of relativistic electrons. 
%
Alternative models, which can also explain the presence of radio mini halos, invoke secondary electrons that are produced by collisions between CR protons and thermal protons  \citep[e.g.,][]{pfrommer04,fujita07,keshet10}. These secondary models are only successful in explaining a small range of observed properties \citep[e.g.,][]{zuhone14}.

\subsection{Is the Phoenix Cluster Unique?}

As one of the most massive cool core clusters known, the Phoenix cluster can be considered, in many ways, a ``scaled-up'', but otherwise normal, system. For example, the $\sim$100\,kpc extent of the star-forming filaments is extreme, being the most extended cool filaments observed in a cool core cluster to date. However, we showed in \cite{mcdonald11b} that multiphase gas is typically observed to a maximum radius of $r_{cool}$ in nearby clusters, where $r_{cool}$ is defined at the radius within which the cooling time is less than 3 Gyr. For Perseus and Abell~1795, this radius corresponds to $\sim$60\,kpc, while for the Phoenix cluster we measure $r_{cool} = 112$\,kpc. Thus, while extreme, the extent of these star-forming filaments are scaling as expected with the overall mass of the cluster, and do not challenge our understanding of cool core clusters. The same can be said for the extreme cooling luminosity, central gas pressure, and radio luminosity.

The extreme star formation rate in the core of the Phoenix cluster ($\sim$600 M$_{\odot}$ yr$^{-1}$) may be pointing to something unique about this cluster. This starburst accounts for $\sim$20\% of the classically-predicted cooling flow, suggesting that cooling may be proceeding very efficiently in this system. For comparison, \cite{odea08} found, for a sample of nearby clusters with star-forming BCGs, a typical ratio of the star formation rate to the cooling rate of 1.8$^{+1.4}_{-0.8}$\% -- roughly an order of magnitude lower than what we observe in the Phoenix cluster. However, when the most extreme end of this population is considered -- clusters harboring starburst BCGs such as Abell~1835 \citep[SFR~$\sim$~200 M$_{\odot}$ yr$^{-1}$;][]{mcnamara06}, RX~J1504.1-0248 \citep[SFR~$\sim$~140 M$_{\odot}$ yr$^{-1}$;][]{ogrean10}, and MACS~1931.8-2634 \citep[SFR~$\sim$~170 M$_{\odot}$ yr$^{-1}$;][]{ehlert11} -- this ratio jumps to $17\pm6$\%. That is, it appears that there is a tail of extreme clusters that are cooling rapidly, fueling efficient star formation in the central galaxy. In this context, the Phoenix cluster represents the most extreme of (but not distinct from) a subset of rapidly-cooling clusters

The cooling properties of the ICM are also, as one may expect, quite different in the Phoenix cluster than in a typical galaxy cluster. In Figure 3 of \cite{voit15}, the normalized density and cooling time profiles of $\sim$100 massive galaxy clusters at $0 < z < 1.2$ are shown. Of these systems, only the Phoenix cluster violates the minimum floor of $t_{cool}/t_{ff} = 10$, and it does so by a substantial margin. The violation of this boundary is thought to initiate strong AGN feedback via condensation or precipitation of the cooling ICM. This is also reflected in Figure \ref{fig:profiles}, where we show that the ICM entropy profile in the Phoenix cluster is consistent with the baseline entropy profile, which represents  the expectation in the absence of any feedback \citep{voit05}, \emph{at all radii}. 
When considering all 165 known galaxy groups and clusters with existing \emph{Chandra} data and temperatures in the range 4 keV $<$ T$_X$ $<$ 15 keV, \cite{cavagnolo09} did not find a single cluster with an entropy profile as steep as Phoenix over the range $10 < r < 1000$ kpc. 
Indeed, the only other cluster known with such a steep entropy profile is H1821+643 \citep{walker14}, which also harbors a cluster-centric QSO, presumably fueled by a rapidly-cooling core. 
The consistency between the entropy profile in the Phoenix cluster and the baseline entropy profile suggests that the ICM has been allowed to cool in this unique system with relatively little resistance.

Perhaps providing the link between these two characteristic aspects of the Phoenix cluster -- the extreme starburst and rapidly-cooling ICM -- is the central AGN, which is unusual in its own way. This central AGN is one of only a few known cluster-centric QSOs \citep{osullivan12,ueda13,kirk15,reynolds14} and appears to be transitioning from quasar-mode to radio-mode. The current radio outburst appears to have recently begun, with the radio-blown bubbles appearing small in size and confined to the innermost regions of the cluster, while a previous outburst may have occurred as recently as $\sim$100\,Myr ago. This AGN appears to be heavily influencing the local environment, highly ionizing the cool gas in the central $\sim$10\,kpc, driving an ionized outflow north of the central galaxy \citep{mcdonald14a}, and inflating bubbles in the dense ICM which may be re-directing the extended, star-forming filaments (see Figure \ref{fig:jets}). Understanding the extreme nature of this AGN, specifically why it appears to have had transitioned from mechanical-mode to radiative-mode, and now back to mechanical-mode, will likely provide the key to understanding the other oddities in the core of the Phoenix cluster.

\section{Summary}

In this paper we present a detailed, multi-wavelength study of the core of the Phoenix galaxy cluster (SPT-CLJ2344-4243). The analysis presented here is based on a combination of new and archival data at radio (GMRT), optical (HST-WFC3, Magellan Megacam), ultraviolet (HST-WFC3, HST-COS), and X-ray (\emph{Chandra}-ACIS-I) wavelengths. The primary results of this latest study are summarized as follows:

\begin{itemize}
\item Complex, star-forming filaments are observed in the rest-frame ultraviolet to extend up to $\sim$100\,kpc from the central cluster galaxy in multiple directions. These newly-detected filaments extend a factor of $\sim$2 times further than the previously-reported maximum extent of star formation.

\item Modeling the combined UV-optical (rest-frame 900\AA--6000\AA) spectrum of the central cluster galaxy reveals a massive ($2.2\times10^9$ M$_{\odot}$), young ($\sim$4.5 Myr) stellar population. Based on these data, we estimate a time-averaged, extinction- and aperture-corrected star formation rate of $610 \pm 50$ M$_{\odot}$ yr$^{-1}$. We note that this estimate can vary by a factor of $\sim$2 in either direction by changing our assumptions on the star formation history.

\item The best-fitting dust model is significantly ``grayer'' than that measured for the Milky Way and SMC/LMC, consistent with observations for nearby starburst galaxies and distant quasars. The data show no evidence of the 2175\AA\ bump, suggesting that there may be a process either destroying or preventing the formation of small grains and those contributing to this bump (i.e., PAHs).

\item We detect significant O\,\textsc{vi}\,$\lambda\lambda$1032,1038 emission (L$_{OVI} = 7.55 \pm 0.20 \times10^{43}$ erg s$^{-1}$) in the inner $\sim$15 kpc of the cluster core. This emission is consistent with having origins primarily in the central AGN and in shock-heated gas along a northern ionized outflow. We are unable to put constraints on what fraction of this emission may originate from the cooling ICM -- the data are consistent with both a complete lack of cooling and a massive (i.e., 5000 M$_{\odot}$ yr$^{-1}$) cooling flow.

\item We confirm the presence of strong (S/N $\sim$ 25) X-ray cavities in the inner 20\,kpc of the cluster core. The total mechanical energy in these cavities is P$_{cav}$ = 2--7~$\times$~10$^{45}$ erg s$^{-1}$, depending on their intrinsic shape, making this one of the most powerful outbursts of radio-mode feedback known. The inferred jet power from these cavities is slightly less than the cooling luminosity (L$_{cool} \sim 10^{46}$ erg s$^{-1}$) in the inner 100 kpc.

\item We find that the bolometric X-ray luminosity of the AGN ($L_{X,bol} = 5.6 \times 10^{45}$ erg s$^{-1}$) corresponds to a few percent of the Eddington luminosity, consistent with a recent transition from quasars to radio-mode AGN. The similarity of the bolometric luminosity and the mechanical power further supports this picture of an AGN transitioning from a radiatively-efficient mode to a mechanical mode.

\item We find evidence for an additional set of X-ray cavities at larger radii ($\sim$100 kpc), suggesting that there may have a been prior episode of radio-mode feedback $\sim$100 Myr ago. Assuming that both potential ``ghost'' cavities are real, this prior episode of feedback may have had jet powers of 2.8 -- 10.5 $\times$ 10$^{45}$ erg s$^{-1}$, which is similar to what is measured in the inner set of cavities. This implies a relatively constant mechanical output of the central AGN between bursts, with a duty cycle of $\sim$100 Myr. 

\item The azimuthally-averaged cooling time of the ICM is shorter than the precipitation threshold ($t_{cool}/t_{ff}$ = 10) at radii of $\lesssim$50 kpc, suggesting that local thermodynamic instabilities in the hot ICM may be fueling both the star formation and AGN feedback. Dense substructures extending beyond 50\,kpc likely have exceeded the precipitation threshold out to much larger radii ($\sim$100-150\,kpc).

\item We see significant substructure in the inner $\sim$200\,kpc (0.15R$_{500}$) of the cluster. The spiral shape of this structure is reminiscent of infalling cool clouds. This may be cooling gas that has been redirected by the strong mechanical feedback, infalling cool group-sized halos, or sloshing of the cool core. The presence of a radio mini-halo supports the X-ray observations that the inner core is highly turbulent.

\item Outside of the cool core ($r>0.15R_{500}$), the cluster appears relaxed and has thermodynamic properties typical of other clusters at similar mass and redshift.

\end{itemize}

While illuminating in many ways, these new data leave several questions unanswered. It remains unclear how such vigorous star formation is sustained in the midst of massive, radio-mode outbursts from the central AGN. The absolute estimate of the star formation rate in the central galaxy hinges on understanding the star formation history, which would benefit from broader-band near-UV spectroscopy, while confirmation that star formation is being fueled by cooling of the hot ICM awaits high resolution X-ray spectroscopy. Confirmation of the extended cavities at large radii requires deeper X-ray follow-up, alongside high angular resolution low-frequency radio imaging. As the most extreme cool core cluster known, such follow-up studies of this systems will allow a deeper understanding of the complex and ongoing war between AGN feedback and cooling in dense cluster cores.
%

\section*{Acknowledgements} 
M. M. acknowledges support by NASA through contracts HST-GO-13456.002A (Hubble) and GO4-15122A (Chandra), and Hubble Fellowship grant HST-HF51308.01-A awarded by the Space Telescope Science Institute, which is operated by the Association of Universities for Research in Astronomy, Inc., for NASA, under contract NAS 5-26555. 
The Guaranteed Time Observations (GTO) included here were selected by the ACIS Instrument Principal Investigator, Gordon P. Garmire, of the Huntingdon Institute for X-ray Astronomy, LLC, which is under contract to the Smithsonian Astrophysical Observatory; Contract SV2-82024.
J.E.C. acknowledges support from National Science Foundation grants PLR-1248097 and PHY-1125897.
B.R.M. acknowledges generous financial support from the Natural Sciences and Engineering Research Council of Canada. 
R.J.W. is supported by NASA through the Einstein Postdoctoral grant number PF2-130104 awarded by the Chandra X-ray Center, which is operated by the Smithsonian Astrophysical Observatory for NASA under contract NAS8-03060.
J.HL is supported by NSERC through the discovery grant and Canada Research Chair programs, as well as FRQNT.
D.A. acknowledges support from the DLR under projects 50 OR 1210 \& 1407, and from the DFG under project AP 253/1-1. 
A.C.E. acknowledges support from STFC grant ST/I001573/1.
%


\end{document}